\documentclass[3p,times]{elsarticle}

\usepackage{ecrc}

\volume{00}

\firstpage{1}

\journalname{Astroparticle Physics}

\jid{procs}

\jnltitlelogo{Astroparticle Physics}

\usepackage{amssymb}

\usepackage[figuresright]{rotating}
\usepackage{url}
\usepackage{color}

\usepackage[normalem]{ulem}

\begin{document}

\begin{frontmatter}




\title{ Preflight performance studies of the PoGOLite hard X-ray polarimeter}


\author[a,b]{M. Chauvin}
\author[a]{M. Jackson}
\author[c]{T. Kawano}
\author[a,b]{M. Kiss}
\author[a,b]{M. Kole}
\author[a,b]{V. Mikhalev}
\author[a,b,d]{E. Moretti\corref{cor1}}
\ead{emoretti@kth.se, moretti@mpp.mpg.de}
\author[c]{H. Takahashi}
\author[a,b]{M. Pearce}


\address[a]{KTH Royal Institute of Technology, Department of Physics, 106 91 Stockholm, Sweden}
\address[b]{The Oskar Klein Centre for Cosmoparticle Physics, AlbaNova University Centre, 106 91 Stockholm, Sweden}
\address[c]{Hiroshima University, Department of Physical Science, Hiroshima 739-8526, Japan}
\address[d]{Max-Planck-Institut f\"ur Physik, D-80805 M\"unchen, Germany}

\cortext[cor1]{Corresponding author, currently at Max-Planck-Institut f\"ur Physik, F\"ohringer ring 6, D-80805 M\"unchen, Germany }

\begin{abstract}

Polarimetric studies of astrophysical sources can make important contributions to resolve the geometry of the emitting region and determine  
the photon emission mechanism. PoGOLite is a balloon-borne polarimeter operating in the hard X-ray band (25-240 keV), 
with a Pathfinder mission focussing on Crab observations.  Within the polarimeter, the distribution of Compton scattering 
angles is used to determine the polarisation fraction and angle of incident photons. To assure an unbiased measurement of the polarisation during a 
balloon flight it is crucial to characterise the performance of the instrument before the launch. This paper presents the results of the PoGOLite 
calibration tests and simulations performed before the 2013 balloon flight. The tests performed confirm that the polarimeter does not have any 
intrinsic asymmetries and therefore does not induce bias into the measurements. Generally, good agreement is found between results from test data 
and simulations which allows the polarimeter performance to be estimated for Crab observations.

\end{abstract}


\end{frontmatter}


\section{Introduction}
\label{introduction}

The development of new observation techniques has enabled enormous progress in X-ray astrophysics as regards spectrometry, imaging and timing 
studies. Polarised X-ray emission is expected to arise from many sources as a result of non-thermal processes in highly asymmetric systems, e.g. 
columns and accretion disks. The addition of polarimetric observations results in two new parameters, the polarisation fraction and angle, and 
would allow geometrical and physical effects to be disentangled, leading to a model independent understanding of underlying emission 
processes~\cite{science1, science2, kamae}. In synchrotron 
processes, the electric field vector of the X-ray flux is perpendicular to the magnetic field lines in the emitting region and hence a polarisation 
measurement determines the direction and the degree of order of the magnetic field at the emission site. Relevant sources are rotation-powered 
neutron stars (e.g. Crab pulsar), pulsar wind 
nebulae (e.g. Crab nebula) and jets in active galactic nuclei (e.g. Markarian 501, 1E1959+65). In Compton scattering processes, the electric 
vector is perpendicular to the plane of scattering and a polarisation measurement determines the geometrical relation between the photon source 
and the scatterer. Accretion disks around black holes (e.g. Cygnus X-1) can be studied in this way. 
There is a paucity of data from instruments specifically designed to make polarimetric measurements (i.e. tested for this purpose prior to flight).
The linear polarisation of X-ray emissions from the Crab were studied over forty years ago 
using a dedicated polarimeter~\cite{weisskopf, weisskopf2}. More recently, inventive use of instruments not originally designed for polarimetric observations 
have reinvigorated the field~\cite{integral, integral2, integral3}.

PoGOLite is a hard X-ray polarimeter which makes observations from a stabilised balloon-borne platform at an 
altitude of approximately 40 km~\cite{kamae}. A reduced effective area polarimeter, the PoGOLite Pathfinder~\cite{Payload_paper}, was first 
launched from the Esrange Space Centre in northern Sweden on 7th July 2011. The balloon was damaged during launch and the flight was terminated 
prematurely, precluding scientific measurements. A successful launch was achieved on 12th July 2013 and resulted in a near-circumpolar flight 
with multiple Crab observations. 

The polarisation of incident X-rays is determined by reconstructing the azimuthal Compton scattering angle ($\eta$) 
in a close-packed array of plastic scintillators with hexagonal cross-section. Tests of the polarimetric concept using a prototype instrument 
exposed to a polarised synchrotron beam have been reported on previously~\cite{beam, beam2, MonteCarlo_paper}. In the current paper, the response 
and performance of the PoGOLite Pathfinder polarimeter was evaluated prior to flight during a series of tests using radioactive 
sources. Polarisation measurements rely on studying counting rate asymmetries in the detector volume and so constitute a positive definite 
quantity. One theme of this work is therefore a study of the polarimeter response to both polarised and unpolarised X-ray beams. A second theme 
is to validate a Geant4~\cite{G4} simulation model of the polarimeter. 

\begin{figure}[!ht]
 \centering
    \includegraphics[width=11 cm]{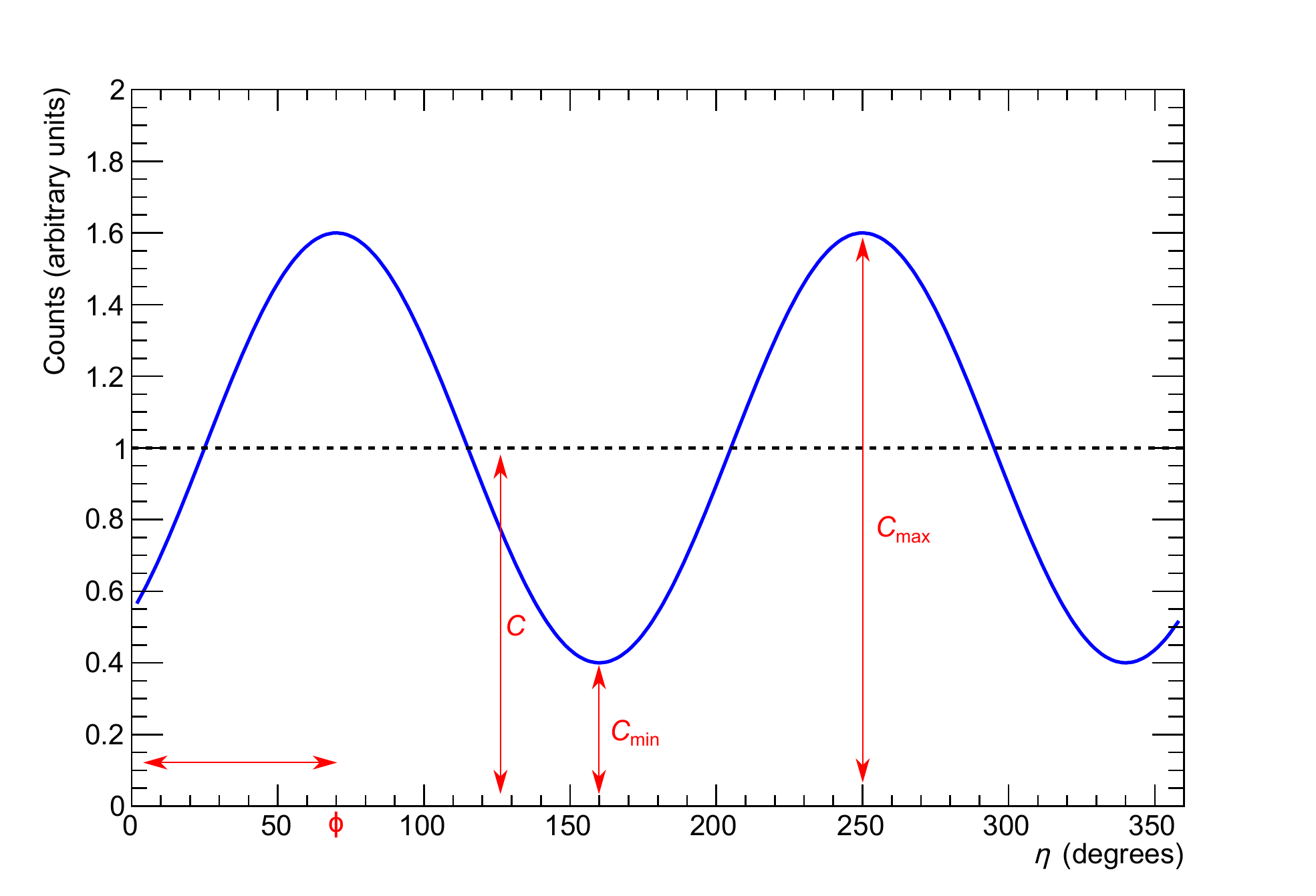}
\caption{A generalised modulation curve, as described by equation \ref{modul_factor}. }
  \label{fig:modcurve}
\end{figure}

An established figure-of-merit for X-ray polarimeters is the 
Minimum Detectable Polarisation ($MDP$)~\cite{MDP},
\begin{equation}
MDP = \frac{4.29}{M_{100} R_s} \sqrt{\frac{R_s + R_b}{T}},
\end{equation}
where $M_{100}$ is the modulation factor (\%) for a 100\% polarised source, $R_s$ ($R_b$) is the signal (background) rate (Hz) for polarisation 
events and $T$ (s) is 
the observation time. From a sinusoidal modulation curve fitted to the distribution of azimuthal scattering angles, the modulation factor, $M$, is 
defined as the ratio between the amplitude and the mean value of the modulation: $M = (C_{max} - C_{min}) / (C_{max} + C_{min})$ ~(see fig.~\ref{fig:modcurve}). 
The factor of 4.29 signifies that this formulation describes the 
capability of rejecting the null hypothesis (no polarisation) at 99\% confidence level. In order to define the MDP corresponding to the Gaussian 
5$\sigma$ detection level, a factor of 7.58 should be used, resulting in a correspondingly longer observation time.
The value of $M_{100}$ is determined from the Geant4 simulation model of the polarimeter. Moreover, the polarisation fraction, $\Pi$, of 
source emission can be expressed as $M / M_{100}$, and the polarisation angle is defined as the phase, $\phi$, of the modulation curve.

For a generalised modulation curve shown in figure~\ref{fig:modcurve} the sinusoidal component with period 180$^{\circ}$ 
arises due to the azimuthal scattering angle dependence of the Compton scattering cross-section, as described by the Klein Nishina equation. The 
modulation curve can be parameterised as   
\begin{equation}
f(\eta) = C \left[1+ M_S \cos \left(\frac{2 \pi}{180} (\eta_S - \phi_S)\right)\right],
\label{modul_factor}
\end{equation}
where, $C = (C_{max} + C_{min}) / 2$. If anisotropic background fluxes are incident on the sensitive volume of the polarimeter, 
two additional sinusoidal components with periods of 180$^{\circ}$ and 360$^{\circ}$ may arise. 
In this case, an appropriate parameterisation is given by   
\begin{equation}
f(\eta) = C \left[1+ M_S \cos\left(\frac{2 \pi}{180} (\eta_S - \phi_S)\right) + M_B \cos\left(\frac{2 \pi}{180} (\eta_B - \phi_B)\right) + 
M_{360} \cos\left(\frac{2 \pi}{360} (\eta - \phi_{360})\right)\right],
\label{modul_factor_180_360}
\end{equation}
where the amplitude and phase of the background contribution are described by $M_B$ and $\phi_B$ in the case of the 180$^{\circ}$ component
and $M_{360}$ and $\phi_{360}$ in the case of the 360$^{\circ}$ component.

In the next section the design of the Pathfinder polarimeter is described. In section~\ref{Acq_trigger} the data acquisition system and trigger 
philosophy are detailed. This is followed in section~\ref{event_sel} by an overview of how candidate polarisation events are selected from data. 
In section~\ref{cal_uniformity}, the calibration procedure for the polarimeter is reviewed. A simulation model of the polarimeter is presented in 
section~\ref{Geant4} and model predictions are compared to experimental data. The polarimetric response to both unpolarised and polarised 
beams is described in section~\ref{inst_pol_rsp}. The impact of the test and simulation results on the scientific performance of the PoGOLite 
Pathfinder mission is discussed in section~\ref{performance}. Conclusions are presented in section~\ref{conclusion}.

\section{Instrument description}
\label{inst_description}

The polarisation of incident photons is determined through the coincident detection of Compton scattering followed by another photon interaction 
like photo-absorption or Compton scattering in a 
close-packed hexagonal array of 61 plastic scintillator elements. The elements in the array, ``phoswich detector 
cells'' (PDCs), each comprise a stack of three scintillators: a \mbox{60 cm} long plastic scintillator tube (Eljen Technology EJ−240), a \mbox{20 cm} 
long plastic scintillator rod (Eljen Technology  EJ−204) and a \mbox{4 cm} long bismuth germanium oxide (BGO) piece. To increase the light 
yield the first two elements of the stack are wrapped by a 3M VM2000 film while the bottom BGO is coated with a layer of $\mathrm{BaSO_4}$-loaded 
epoxy resin. Each PDC has a hexagonal cross-section of about \mbox{30 mm}.
The three elements provide active collimation, an active target for scattering or absorption 
and bottom anticoincidence, respectively. Each PDC is read out by a photomultiplier tube (PMT), Hamamatsu Photonics~\cite{Hamamatsu} 
R7899EGKNP. To provide additional collimation, each hollow plastic tube is first wrapped in a tin foil and then a lead one, each 50~$\mu$m thick. The lead 
foil absorbs off-axis photons and the tin foil absorbs fluorescence photons emitted by the lead. The hollow scintillator tube and BGO piece have decay 
times of about 300 ns (``slow''), while the solid scintillator rod has a decay time of 2 ns (``fast'').
Since the component materials have different scintillation decay times, pulse shape discrimination allows vetoing events with interactions 
in the slow plastic scintillator or the BGO. 

\begin{figure}[!ht]
 \centering
\includegraphics[width=11 cm]{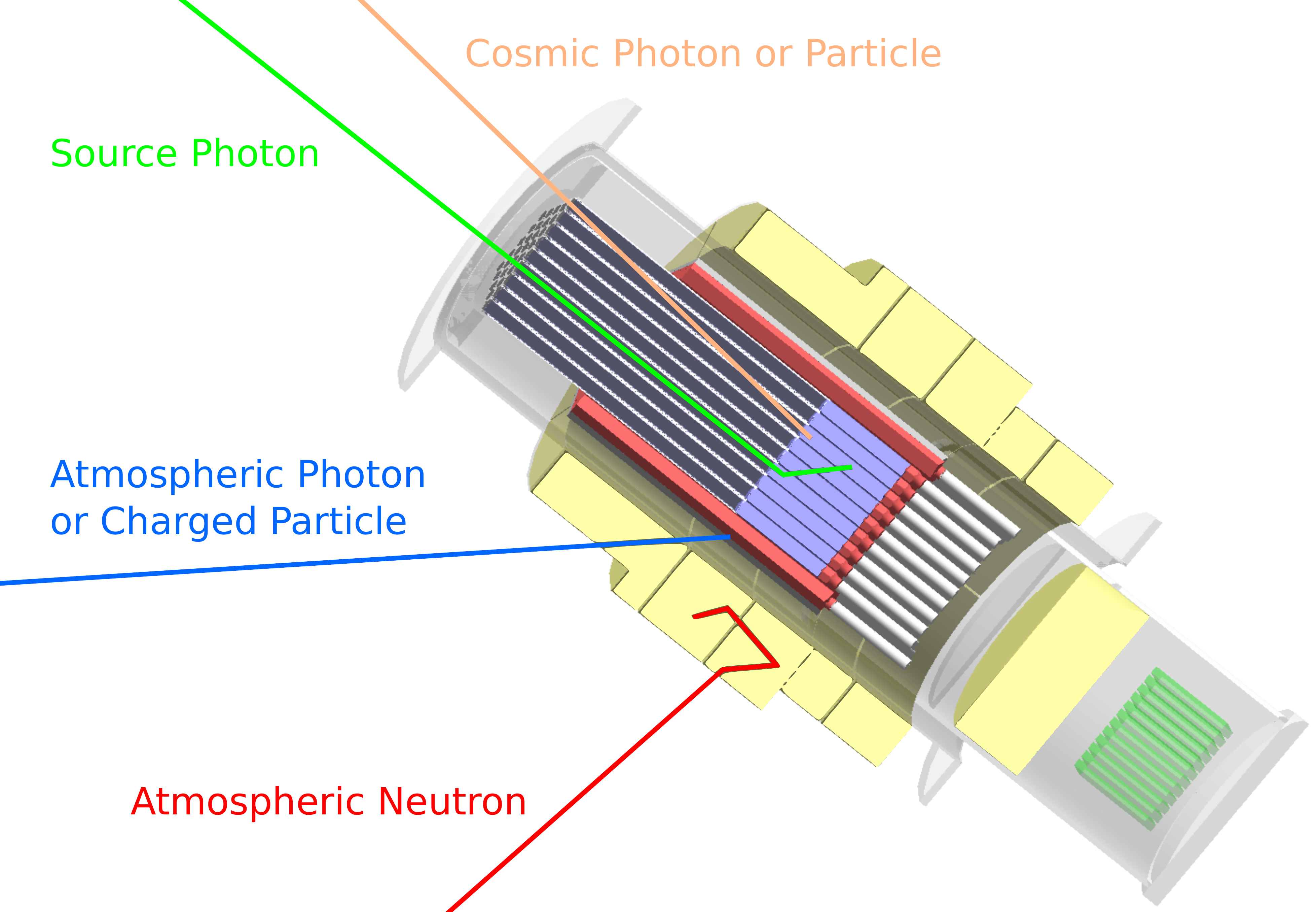}
\caption{A schematic cross-section of the PoGOLite Pathfinder instrument. The scintillator scattering target is shown in blue, while
the hollow collimating scintillators are shown in black and the BGO units in red. The scintillators are read out by photomultipliers (white). 
The polarimeter is housed in a pressure vessel (grey), which in turn is surrounded by a polyethylene neutron shield (yellow). The data 
acquisition electronics (green) are placed 
at the bottom of the instrument. Possible interactions of photons and particles with the detector elements are indicated. 
The telescope measures about 1.5 m from top to bottom and its aperture is about 30 cm wide.}
\label{fig:Detector_array_Geant}
\end{figure}

The detector array is surrounded by a segmented active side anticoincidence shield (SAS) made of BGO. This provides shielding against 
interactions from charged particles and photons originating from outside the instrument field-of-view, which is  
$2.4^{\circ} \times 2.6^{\circ}$, 
defined by the geometry of the plastic scintillator collimation tubes. SAS elements are read out by the same type of photomultiplier tubes 
as used for the PDCs, and the segmentation of the shield allows anisotropies in the background environment to be studied. 
A passive polyethylene shield with a thickness varying between \mbox{10 cm} and \mbox{15 cm} surrounds the detector array and reduces 
background from atmospheric neutrons, the dominant measurement background  during balloon flights \cite{merlinPhD}. 

The detector array, photomultiplier tubes and associated electronics are housed inside a pressure vessel, which is rotated around the 
instrument viewing axis in order to cancel out systematic effects during measurements. These can arise due to intrinsic differences between 
PDC response or if a photomultiplier tube fails (as is the case for the measurements presented in this paper).
One full revolution is completed during a standard 5 minute long data acquisition run. The detector array is shown in 
figure~\ref{fig:Detector_array_Geant}. 

An attitude control system~\cite{DST} allows the instrument to acquire and track observational targets 
on the sky. It combines inputs from various sensors including differential GPS, magnetometers and optical star trackers and produces a 
pointing solution and corresponding control feedback for the actuator motors that govern the azimuth and elevation pointing of the instrument. 
The GPS is also used to provide an absolute time reference for photon time-tagging. 

\section{Data acquisition and trigger}
\label{Acq_trigger}

The PoGOLite data acquisition system is built around analogue front-end electronics and waveform digitisers (implemented on 12 Flash Analogue to 
Digital Converter (FADC) boards), trigger/event logic (implemented on a digital input/output (DIO) board) and a control computer. Inter-board 
communication is through the SpaceWire standard. This architecture is described in more detail elsewhere~\cite{spie_ht_2014}. The trigger system 
allows events to be selected based on the photomultiplier pulse shapes and veto information 
from the side anticoincidence system. Event selections are defined in order to distinguish background events from valid Compton scattering and 
photo-electric interactions in the fast plastic scintillators. At trigger level, relatively conservative selections are made in order to avoid 
bias.  More stringent selections are made offline once data are downlinked to the ground.

\begin{figure}[!ht]
 \centering
\includegraphics[width=0.9\textwidth]{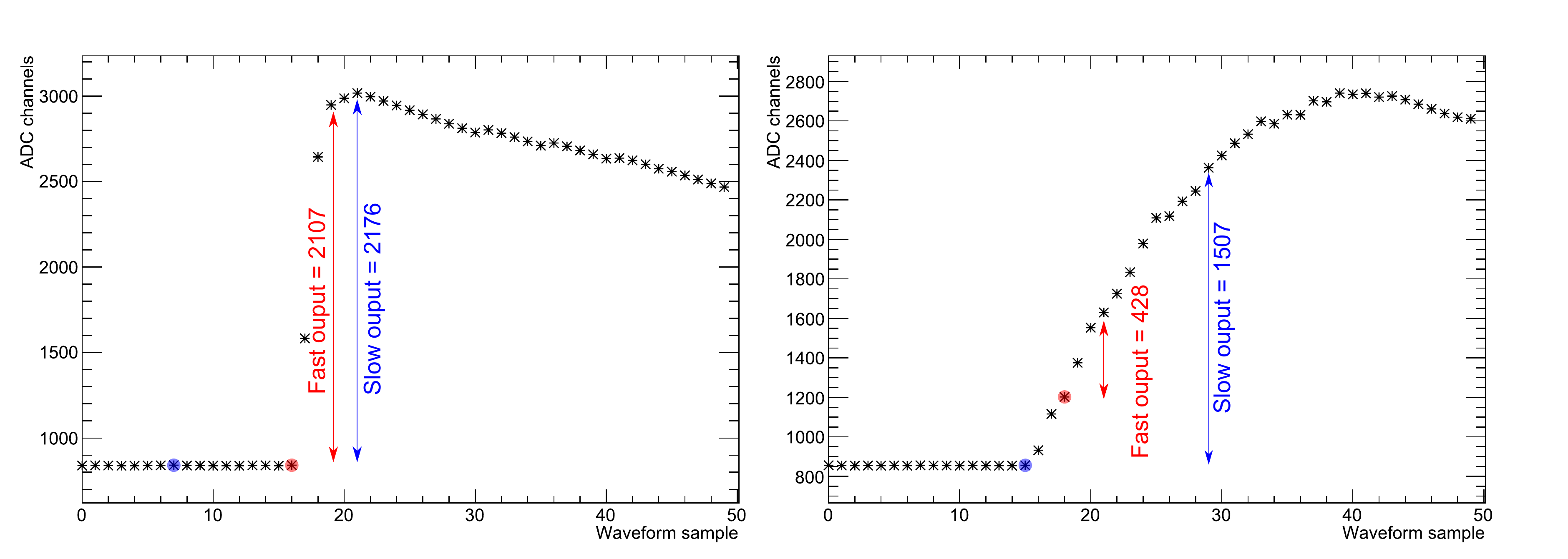}
\caption{An example of interactions on the fast (left plot) and slow (right plot) scintillators and their corresponding fast and slow outputs as calculated 
during online selection. For the fast scintillator the outputs are similar in magnitude. For the slow scintillator the maximum slow output is much 
greater than the maximum
fast output. The online waveform discriminator logic relies on the difference between the fast and slow outputs in the case of a fast or slow waveform. } 

\label{fig:waveforms}
\end{figure}

Signals from the last dynodes of all PMTs are fed to individual FADC channels and digitised with 12 bit accuracy at a 37.5 MHz sampling rate. 
Field-Programmable Gate Arrays (FPGA) on the FADC boards continuously monitor the waveforms and calculate the pulse height for each sample point.
Pulse heights are corrected for the baseline by subtracting the ADC value measured 3 clocks before. 
A 'hit' signal is issued by the trigger system for a given photomultiplier when the pulse height is greater than $\sim$0.5 keV (10 ADC 
channels). This threshold value was chosen to avoid electronic noise and to match the dark current properties of the photomultipliers.
If the pulse height exceeds $\sim$15 keV (300 ADC channels) a 'level-0 trigger' (L0) signal is issued.
 Given the energy resolution of plastic scintillator, this allows the photo-absorption of photons with an initial energy of $\sim$25 keV 
 to result in a trigger.
An Upper Discrimination (UD) level defines a maximum reconstructed energy of $\sim$120 keV (2500 ADC) per single hit. 
For the side anticoincidence 
detectors the energy threshold is $\sim$15 keV (10 ADC).

\begin{figure}[!ht]
 \centering
\includegraphics[width=11 cm]{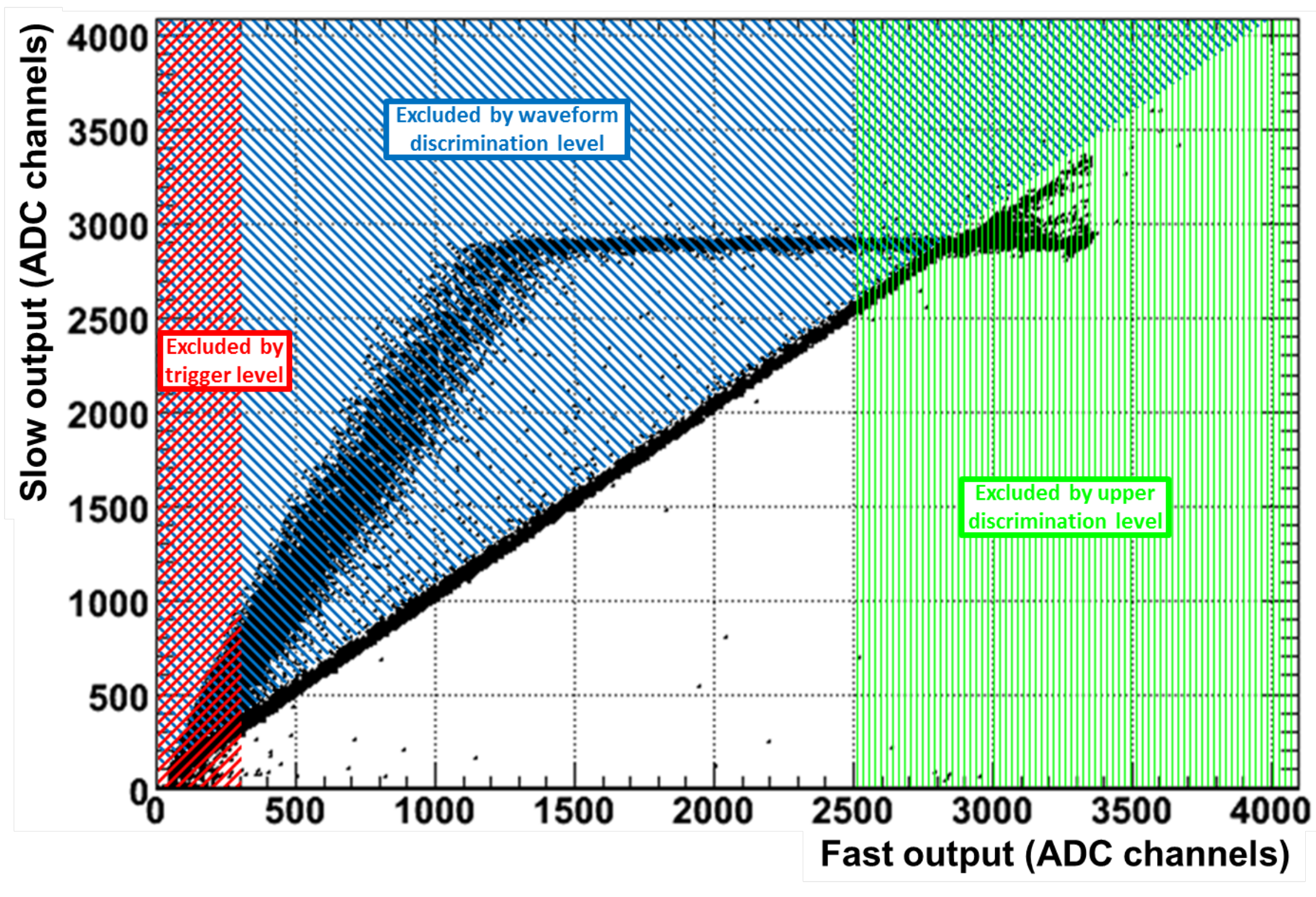}
\caption{Photons interacting in the slow scintillator tube or bottom BGO piece (see section~\ref{inst_description}) can be 
rejected in the data acquisition system using various trigger and threshold settings. The trigger level (red) and upper discrimination (UD)
threshold (green) discard noise events and saturated events. The waveform discrimination (WD) threshold (blue) is used to 
remove events with interactions in the slow or bottom BGO scintillators. The resulting acceptance region corresponds to clean hits which are 
used for the polarisation analysis.}
\label{fig:veto_Fast_slow}
\end{figure}

Pulse shape discrimination at trigger level is performed by comparing the maximum pulse heights measured with 3 clock cycles ($\sim$80ns, 
fast output) and with 14 clock cycles separation ($\sim$370ns, slow output), as shown in figure \ref{fig:waveforms}. Signals coming from the fast 
scintillators generate fast and slow output signals with comparable pulse heights. Signals coming from the slow or BGO 
scintillators have a smaller fast output signal with respect to their slow output signal. If the pulse falls in the region of slow waveforms 
(blue shaded area of figure \ref{fig:veto_Fast_slow}) a WD (Waveform Discriminator) signal is issued. 
Hits coming from the SAS units issue only WD or UD signals and are used exclusively to veto events. 

When the veto is active, the global event logic implemented on the DIO board collects all L0 triggers and vetoes from the PDC and SAS units and 
issues an acquisition request (L1 trigger) if no UD (green shaded area of figure \ref{fig:veto_Fast_slow}) or 
WD signals have been issued. The acquisition request is distributed to all FADC boards. 
When an FADC board receives an acquisition request, it stores 15 pre-trigger and 35 post-trigger sample points after zero suppression. 
The pre-trigger samples are used to correct for possible baseline shift due to preceding large signals or to identify photomultiplier after-pulses 
by their initial falling slope behaviour. 

A photon Compton scattering in one fast scintillator followed by an interaction in another fast scintillator defines the scattering angle. 
For this reason, the results presented in this paper derive exclusively from two hit events. 
The lower (higher) energy hit is assumed to correspond to the Compton scattering (photo-absorption) interaction. An energy deposit above 300 ADC 
triggers the data acquisition.
From a collimated beam illuminating the central PDC, the fraction of two hit events respect to all the recorded events is about 32\%, the fraction  
of 1 hit events is about 28\%, while the fraction of 3 hit events is 22\%. Finally the fraction of the events with more than 3 hits is 18\%. 

\subsection{Online selection performance}

The online selection system comprises a fast/slow pulse shape discriminator to veto out-of-acceptance events and an
anticoincidence system. Online selections reduce the amount of data stored, thereby decreasing the instrument dead-time 
by discarding background events while keeping most of the signal. Stored events can be further processed during the offline analysis phase to 
improve the signal-to-background ratio. 

\begin{figure}[!ht]
 \centering
    \includegraphics[width=11 cm]{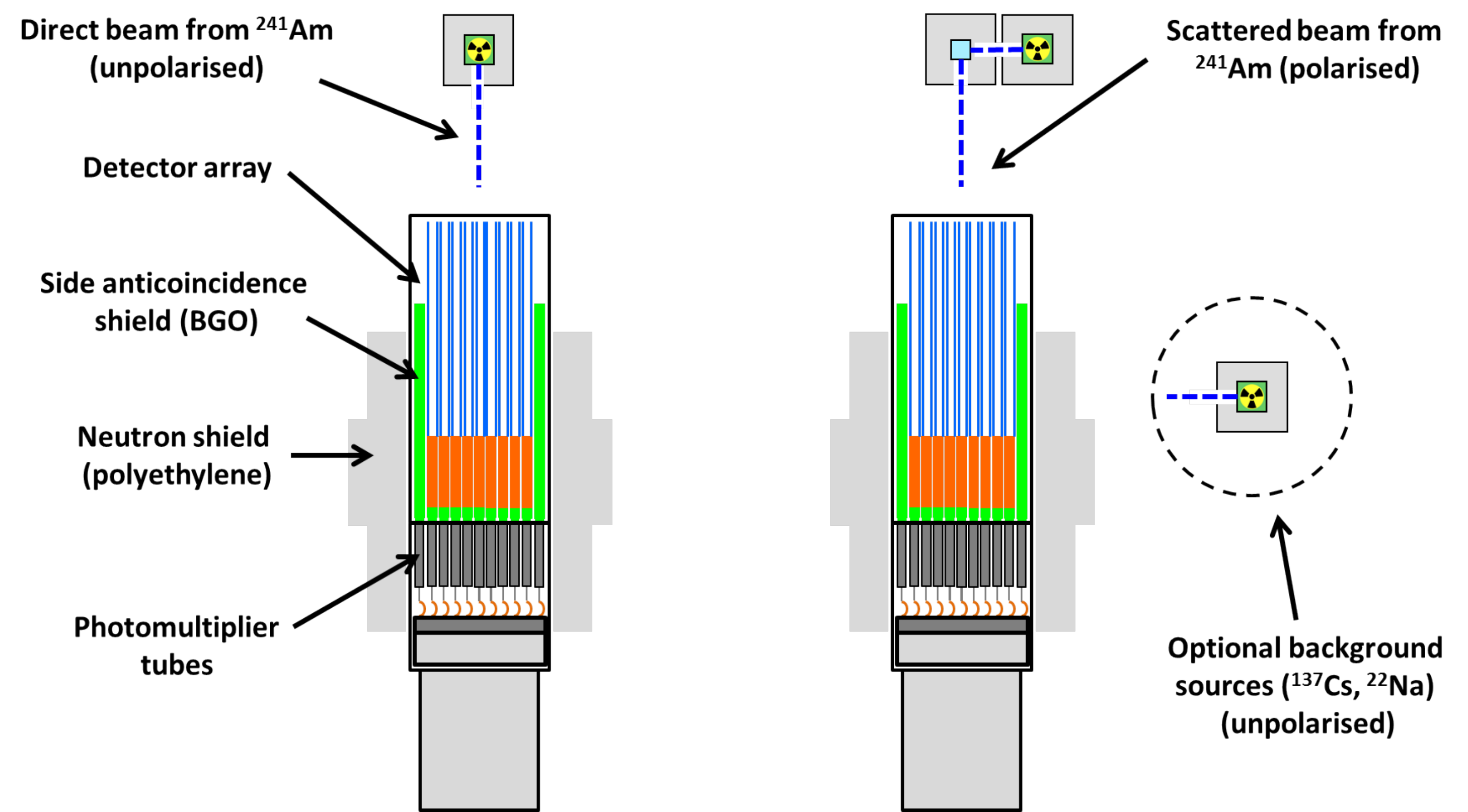}
  \caption{Two experimental set-ups used for the polarimeter tests presented in this paper. In the set-up on the left a collimated $^{241}$Am 
  source illuminates the central PDC of the polarimeter. In this configuration the photon beam is unpolarised. In the set-up on the right the 
  $^{241}$Am source beam is scattered on a polyethylene piece and then illuminates the central PDC of the instrument. In this configuration the 
  photon beam is polarised. In some tests two extra sources ($^{137}$Cs and $^{22}$Na) have been added on the side of the polarimeter.}
     \label{fig:setups}
\end{figure} 

\begin{figure}[!ht]
 \centering
  \includegraphics[width=0.8\textwidth]{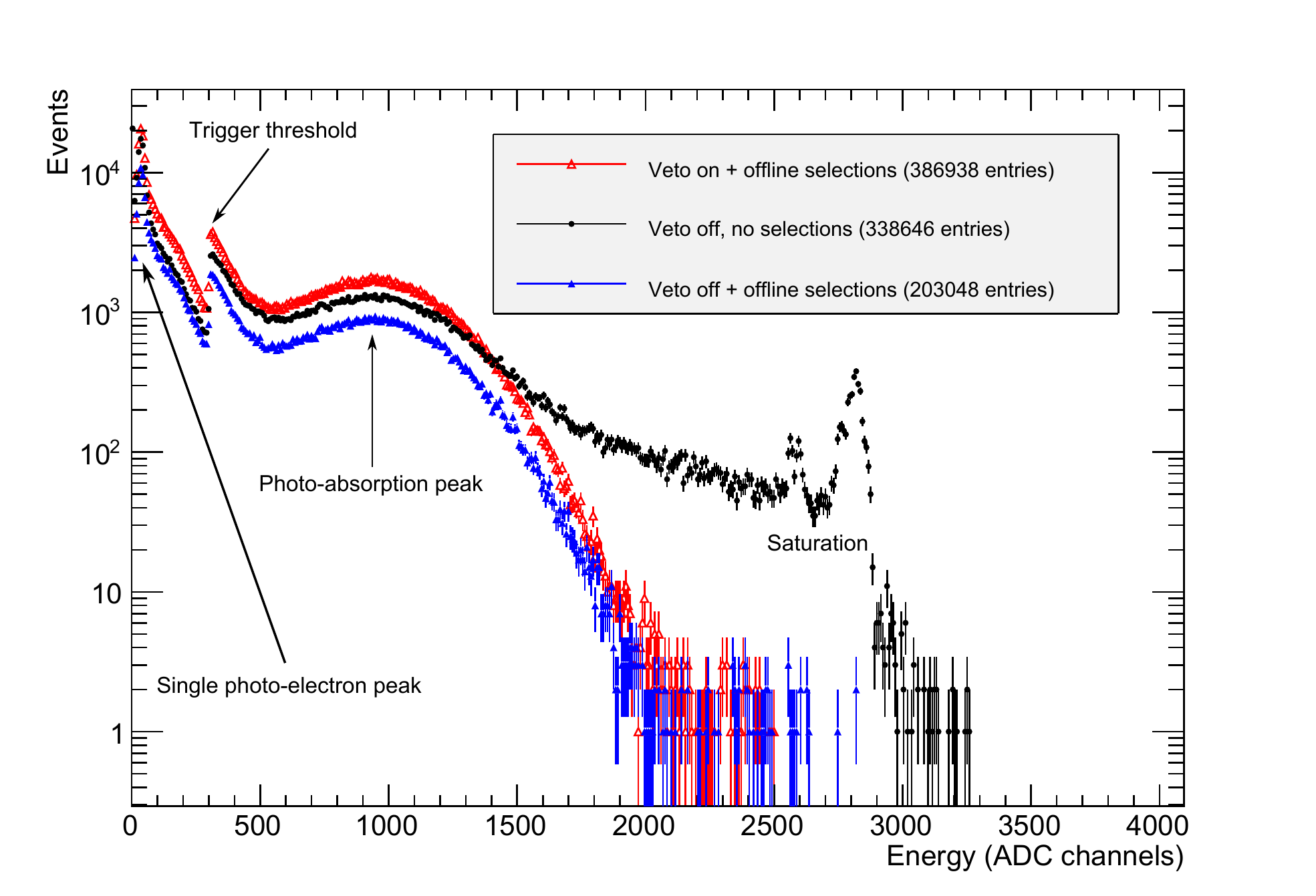}
  \caption{Signal selection performances of the online selection system (veto) in the case of two hit events coming from a collimated $^{241}$Am source 
	   illuminating the central PDC. Each histogram entry corresponds to an interaction in a two-hit event. Entries below the trigger threshold 
	   result from interactions where the data acquisition has been triggered by the other hit. 
	   The black (filled dots) and the blue (filled triangle) histograms refer to a 5 minute measurement with no online 
           selections (veto off), but with signal selection requirements applied offline in the case of the blue (filled triangle) histogram. The 
           red (open triangle) histogram refers, instead, to a 
           5 minute measurement with online selections (veto on) and offline selections. The online veto system does not modify the spectrum of 
           the incoming photon beam, but it increases the amount of collected signal photons. The high energy part of the black full dots spectrum 
           shows two peaked structures formed by saturated events from charged particles.}
    \label{fig:veto_sel_signal}
\end{figure}

To evaluate the performance of online selections, a high flux (186 Hz two hit event rate induced in the polarimeter) collimated $^{241}$Am source 
was used. The set-up is shown on the left side of figure~\ref{fig:setups}. Measurements focused on studies of possible  
spectral bias and the linearity of the reconstructed count rate.
The $^{241}$Am source is enclosed in a stainless steel container to suppress low energy X-rays, resulting in a beam of predominantly 
59.5 keV photons. The source and the container are housed in a block of lead to allow safe handling and rough collimation (opening angle 
{$\pm14^{\circ}$}). 
The photon beam can be further collimated using additional lead blocks with a 2~mm diameter aperture.
Figure \ref{fig:veto_sel_signal} shows the reconstructed energy spectrum for the fully collimated source illuminating the central PDC.

The three spectra shown in figure ~\ref{fig:veto_sel_signal} are populated by two hit events collected during 5 minute long
measurements and show a photo-absorption peak at about 1000 ADC (53.3 keV). Below 500 ADC, in the Compton region, the spectra show the trigger 
threshold at 300 ADC (main hit) and the single photon-electron peak of the hits just over the 10 ADC threshold (second hit). 
The black histogram (filled dots) is created without requiring any online or offline (see sec.~\ref{event_sel}) selections. It is noted 
that although the source rate is much higher than the background rate, a high energy background is present in this histogram.
The blue histogram (filled triangles) contains only those events that passed the offline selections. 
The red histogram (open triangles) shows events that passed both the online and offline event selections. 
Comparing these histograms shows that the online selections do not modify the signal spectrum significantly and demonstrates that removal of 
background events allows a higher event rate to be recorded.
The online and offline selection system has also been tested in an environment dominated by background. The results are presented
in section~\ref{event_sel} and, in particular, figure~\ref{fig:veto_sel_signalVSside_BKG}.   

\subsection{Dead-time}

\begin{figure}[!ht]
 \centering
    \includegraphics[width=11 cm]{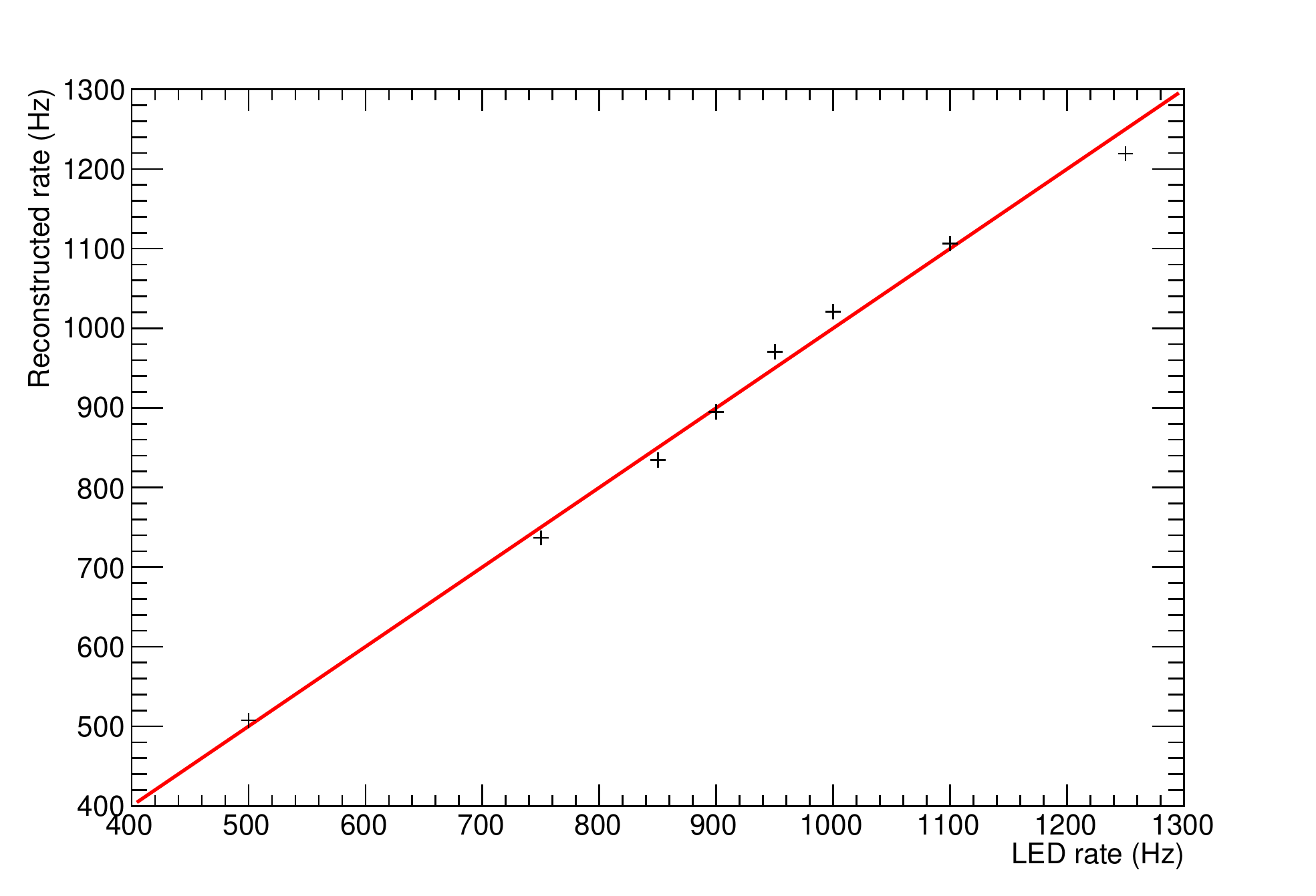}
  \caption{The reconstructed signal rate as a function of the input signal rate (created by a flashing LED) superimposed with 
  the function $f(x)=x$. The signal rate can be reconstructed correctly up to at least 1.25 kHz.}
     \label{fig:reconstructed_rate}
\end{figure}

The reconstructed signal rate must be corrected for measurement dead-time. 
Incorrect reconstruction will induce an error on the inferred signal-to-background ratio during observations which ultimately 
results in an error on the reconstructed polarisation. 
A test was devised whereby the reconstruction of a known signal rate was studied. 
A pulse generator driven LED was placed on top of the central PDC and used to define the input signal rate.
The LED pulse length was set to 10 ns to produce a fast pulse shape from the PDC.
The LED repetition rate was varied between 500~Hz and 1.25~kHz and 
the reconstructed signal rate is shown in figure \ref{fig:reconstructed_rate}.
The signal rate is reconstructed correctly up to 1.25~kHz. At this point, the instrument dead-time exceeds $65\%$.
Experience from the 2011 flight indicates that the signal rates are significantly smaller than 1 kHz.

\subsection{Time tagging accuracy}

A pulse per second (PPS) signal from the GPS unit in the attitude control system indicates the start of each second with high precision. 
Using this to synchronise absolute GPS timing information and combining with a signal from a calibrated oscillator on each FADC board in the 
polarimeter data acquisition system allows an absolute $\mu$s precision time stamp to be associated with polarisation events. The Crab pulsar 
light curve can then be reconstructed once barycentring corrections are applied.

\begin{figure}[!ht]
 \centering
    \includegraphics[width=11 cm]{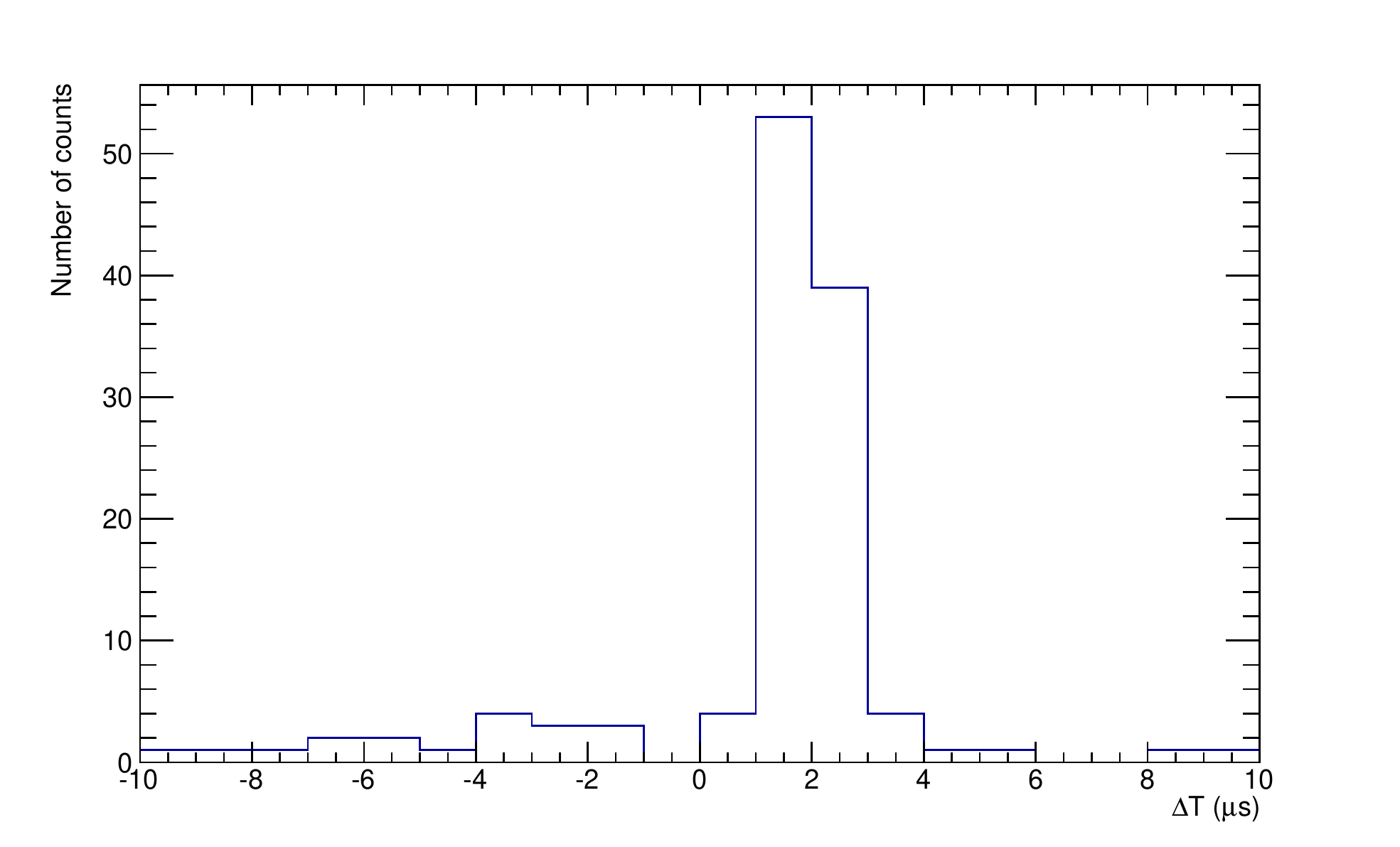}
\caption{The time difference between cosmic-ray muons traversing the polarimeter as registered by the PoGOLite data acquisition system and 
an external scintillator detector.}
  \label{fig:PPS}
\end{figure}

The accuracy of the polarimeter time tagging system was measured using an external scintillator detector for cosmic-ray muons \cite{merlinPhD}. 
The detector was positioned to allow muons which traversed the polarimeter (positioned horizontally) to be identified. The resulting trigger 
signal was fed to a purpose-built FPGA board which used signals from a GPS receiver and a 100 MHz internal oscillator to determine an absolute time for 
the muon trigger.
The distribution of muon crossing times registered by the scintillator detector subtracted from the time reconstructed by the PoGOLite data 
acquisition system, $\Delta$T, is shown in figure \ref{fig:PPS}. 

The value of $\Delta T$ is seen to peak at $\sim$1 $\mu s$, indicating that PoGOLite is 
$\sim$1 $\mathrm{\mu s}$ slower in registering a muon crossing event than the external scintillator detector. 
The width of the peak defines the absolute timing precision to be of order 1~$\mathrm{\mu s}$, thereby fulfilling the requirement for Crab light 
curve reconstruction.

\section{Offline event selection}
\label{event_sel}

An online veto system is used to pre-select signal-like events. Since there is limited online computation time available, a more 
accurate selection is carried out offline. The offline logic to distinguish between signal and background is based on the fast waveform discrimination
and on the temporal coincidence of two hits. By selecting the fast waveforms the events that interact only with the fast scintillator are 
selected. Side entering particles and out-of-acceptance photons that interact with the slow shaping scintillators are therefore discarded.
The following offline selections are applied to stored waveforms: (i) a check on the waveform behaviour after the first peak to remove pulses 
that have multiple peaks, (ii) a falling slope check, on the pre-rising sample point, to remove PMT after-pulses, and (iii) a check on the number 
of clock cycles in the rising 
part of the waveform to remove residual slow waveforms (no more than 6 clock cycles are allowed). After the offline selection of the hits on the 
basis of their waveforms, the time correlation between the main trigger and other hits is considered. The waveform for accepted hits must start to 
rise within $\pm 2$ clock cycles (about 60 ns) from the onset of the rising slope of the trigger waveform. 

\begin{figure}[!ht]
 \centering
  \includegraphics[width=11 cm]{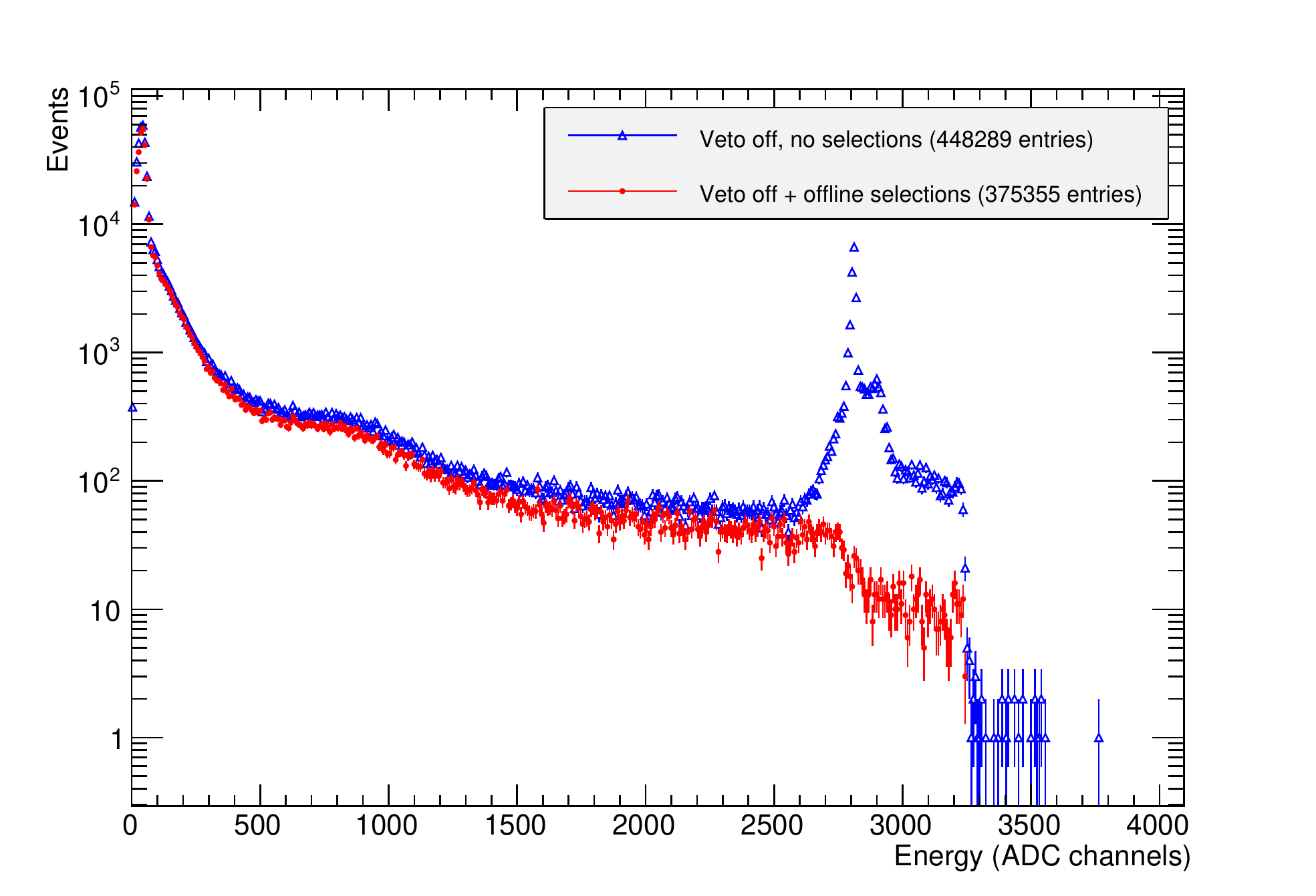}
  \caption{The effect of offline event selections for the case where waveforms come only from the fast scintillator. Data was collected without 
  online selections (veto off). 
  The selection efficiency seems to decrease above 2500 ADC channels due to the saturated waveforms present in the recorded spectrum (blue, open
  triangles) that are discarded from the offline selections.}
    \label{fig:fast_selection}
\end{figure}

\begin{figure}[!ht]
 \centering
  \includegraphics[width=11 cm]{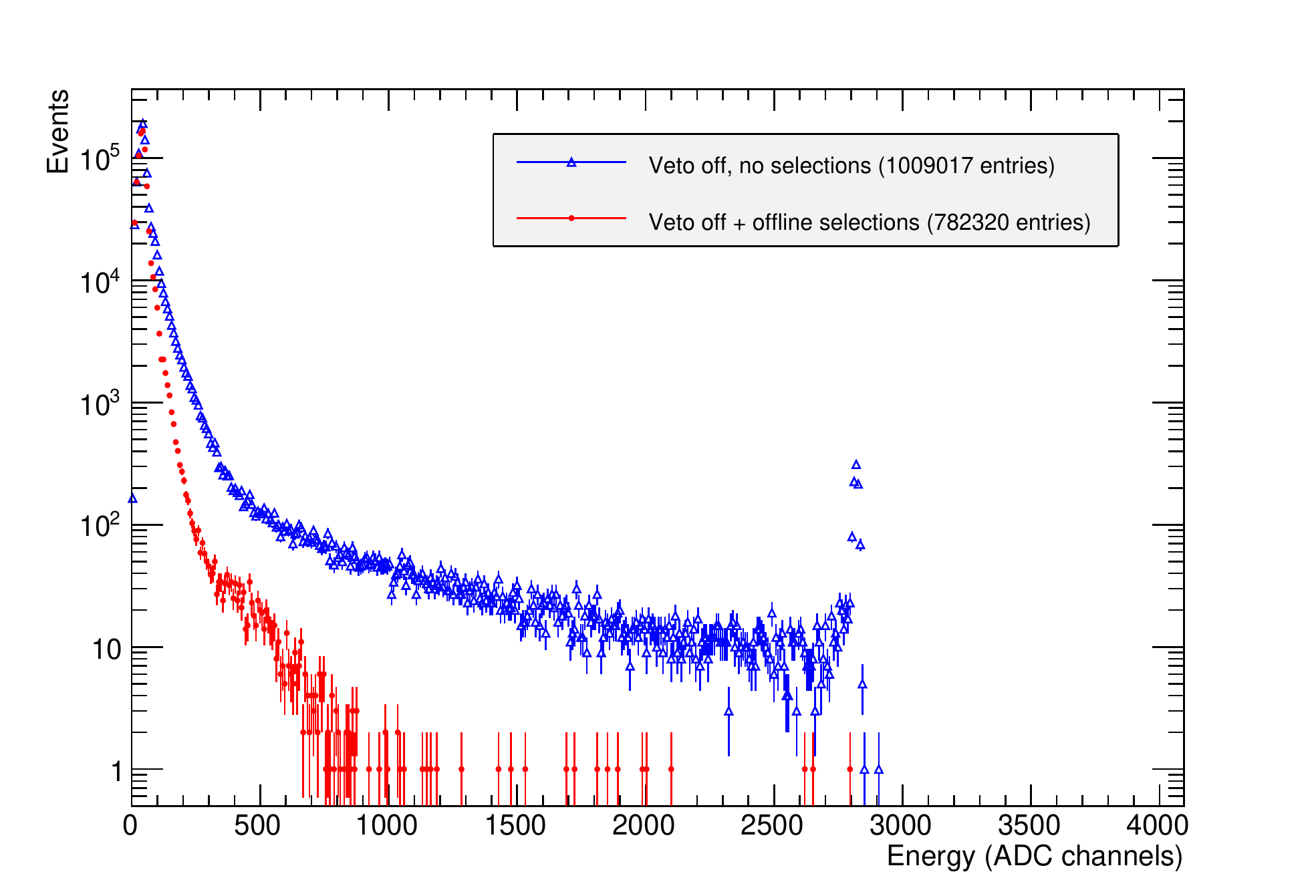}
  \caption{The effect of offline event selections for the case where waveforms come only from the hollow slow scintillator. The data was collected 
  without online selections (veto off). The efficiency of the selections in the trigger region (above 300 ADC) is much higher than in the 
  low energy part of the spectrum.}
    \label{fig:slow_selection}
\end{figure}

\begin{figure}[!ht]
 \centering
  \includegraphics[width=11 cm]{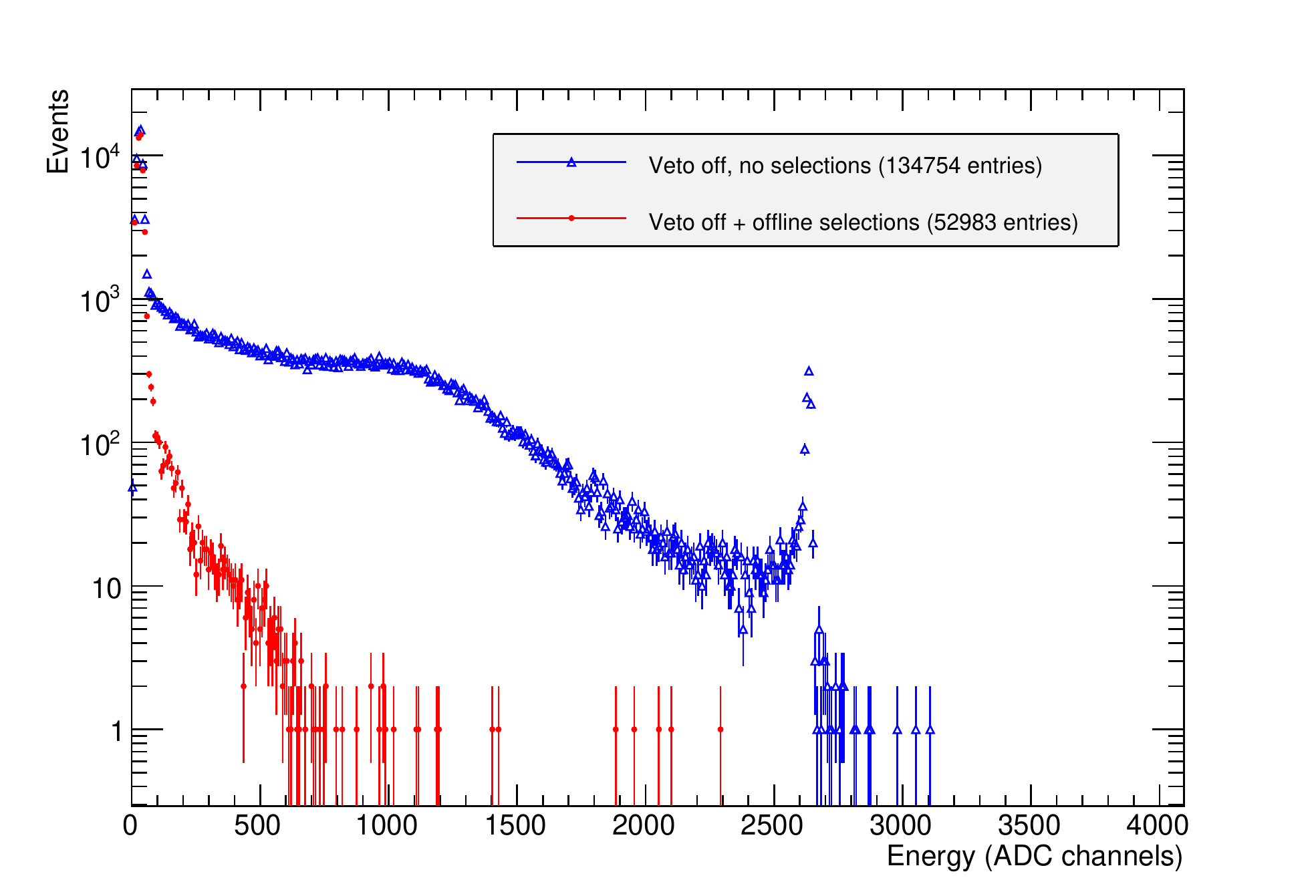}
  \caption{The effect of offline event selections for the case where waveforms come only from the slow bottom BGO scintillator. The data was 
  collected without online selections (veto off). The efficiency of the selections in the trigger region (above 300 ADC) is much higher than in 
  the low energy part of the spectrum.}
   \label{fig:BGO_selection}
\end{figure}

The performance of the offline selections in discriminating between slow and fast waveforms can be determined using an external test set-up 
where only one kind of scintillator is attached to a photomultiplier at a time. The photomultiplier is read out using the PoGOLite data acquisition 
system. 
Since the recorded data sets will only contain waveforms coming from a specific type of scintillator, the efficiency to accept or reject that type 
of waveform can be quantified and therefore the acceptance or rejection of signal and background waveforms can be determined. 
The efficiency can also be studied as a function of the peak ADC value of the waveform, to distinguish the effects of selecting trigger (pulses 
with the maximum above 300 ADC) or hit (pulses with the maximum above 10 ADC) waveforms. 
The acceptance for fast signals (both hits and triggers) can be seen in figure~\ref{fig:fast_selection} to be about $90\%$ up to 2500 ADC channels. 
For slow BGO waveforms (see figure~\ref{fig:BGO_selection}) the contamination of a trigger 
waveform is less than $1\%$ and the contamination of a hit waveform is about $70\%$. For slow waveforms coming from the hollow slow scintillator 
(figure~\ref{fig:slow_selection}), the rejection is about $93\%$ above the trigger threshold and $20\%$ below the trigger 
threshold.

Overall the offline selection has a high acceptance for fast signal hits ($90\%$), while approximately 
$5\%$ of the pulses coming from scintillators with a slow shaping time erroneously result in valid trigger signal.

\begin{figure}[!ht]
 \centering
\includegraphics[width=15 cm]{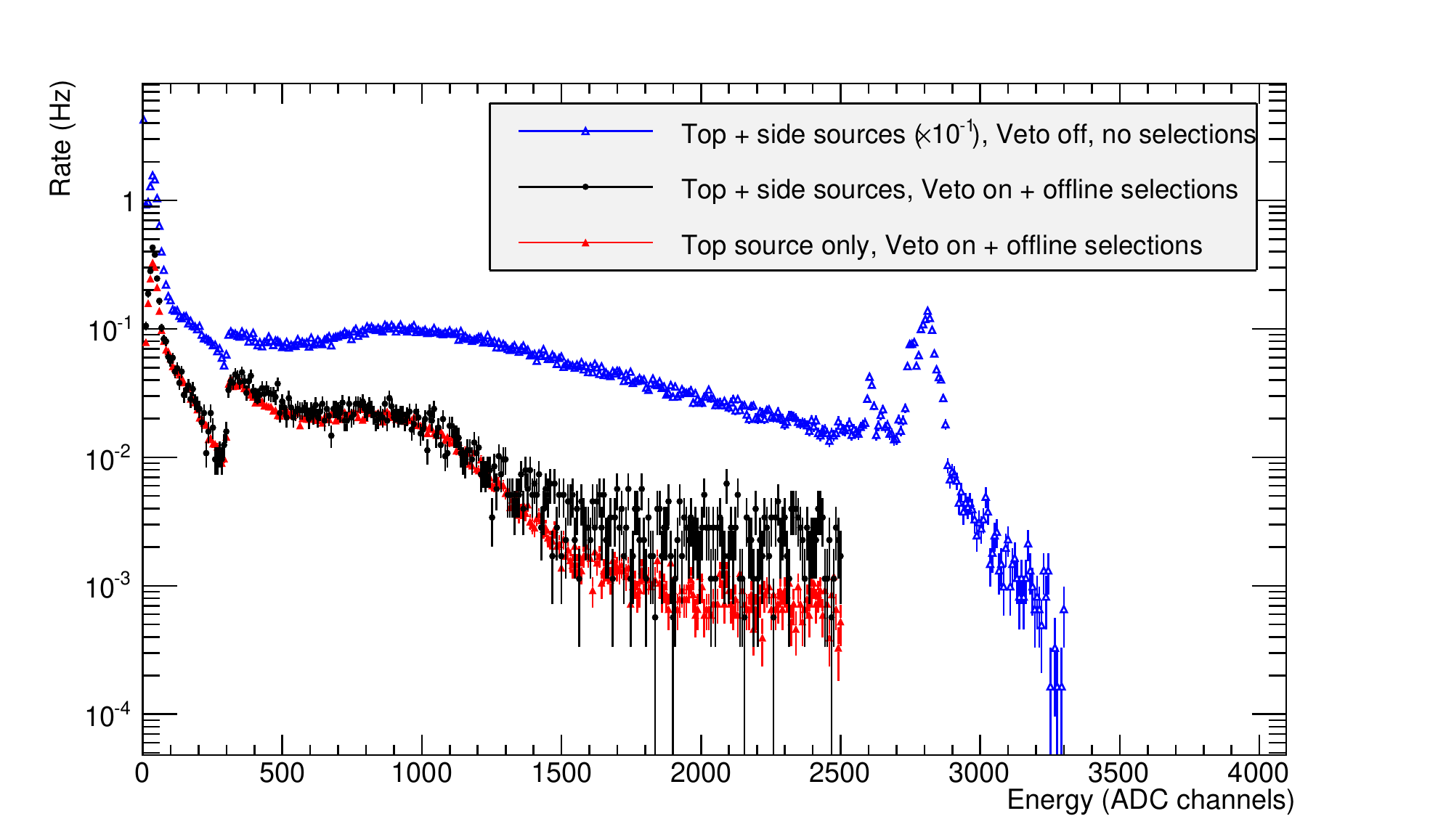}
\caption{Online (black histogram, filled dots) and offline (blue histogram, filled triangles) selection performances in rejecting the events 
coming from a $^{137}$Cs source and a $^{22}$Na source placed on the 
side of the instrument and illuminating the SAS anticoincidence detectors. The on-axis signal photon beam (placed on top of the polarimeter) was 
attenuated by means of a scatter piece while the background was increased using the two lateral sources. The resulting spectrum (black, filled 
dots) compares well with the spectrum from the signal photon beam only (red histogram, filled triangles), confirming the capabilities of the event 
selections.}
\label{fig:veto_sel_signalVSside_BKG}
\end{figure}

The results from a study of the performance of both the online and the offline selections are presented in 
figure~\ref{fig:veto_sel_signalVSside_BKG}. 
The signal rate was decreased by scattering source photons through $90^{\circ}$ with a polyethylene target. The exit aperture has an opening angle 
of approximately $\pm63^{\circ}$ (see the set-up on the right of figure~\ref{fig:setups}). 
The scattered photon beam (53.3 keV) illuminates the central PDC and two hit events are recorded.
A flux of lateral background particles is created using 
$^{137}$Cs and $^{22}$Na sources positioned next to the SAS anticoincidence scintillators and illuminating them orthogonally 
with respect the axis of the polarimeter (see the set-up on the right of the figure~\ref{fig:setups}). 
The lateral sources induce a hit rate in the SAS scintillators of 4.1 kHz, comparable to that induced by background during the 2011 flight.
The blue histogram (filled triangles) contains two hit events recorded without any selections, while two hit 
events that fulfil the online and offline selections form the black (filled dots) histogram. 
The performance of the selections can be evaluated by comparing the black histogram with the red (filled triangle) one, for which there is no 
lateral background. The large number of background hits arising from the lateral sources is efficiently discarded by the online and offline 
selections, although some residual events at high energy penetrate the SAS anticoincidence shield without rejection. 
The spectrum from the signal source is correctly reproduced for the different set-ups and the reconstructed flux agrees within 15\%. 

\section{Instrument calibration and uniformity}
\label{cal_uniformity}

The PDCs are calibrated using the collimated $^{241}$Am source. Each PDC aperture is irradiated in turn, and the position of the \mbox{59.5 keV} 
photo-absorption peak is used to provide a linear energy calibration in ADC channels. For low energy depositions, 
the scintillator light-yield is non-linear and a correction is applied based on previous tests where the same type of plastic 
scintillator was irradiated using synchrotron photons with energies between \mbox{6 keV} and \mbox{70 keV}~\cite{MonteCarlo_paper}. The 
single photo-electron position is also determined for each PDC providing a PMT gain calibration. Photomultiplier tube 
voltages are set such that the photo-peak (59.5 keV) is approximately in the same position in the spectrum (same pulse height) for all units, 
thus providing a reasonably uniform detector response. Any remaining unit-to-unit differences are removed through the rotation of 
the detector array relative to the instrument viewing axis. 

During the calibration procedures, the PDC detection efficiency was also tested. A highly collimated photon beam from an $^{241}$Am source
was co-aligned with the centre of the central PDC. The detection efficiency was calculated from the rate of detected 
two hit only events (normalised by the live-time of the detector) divided by the incoming photon rate. The latter rate value was cross checked in 
three different ways: (i) by measuring the same collimated source with another more efficient apparatus, (ii) by simulating the experimental 
set-up, and (iii) with an analytical calculation that took into account the nominal source activity and the geometry of the collimation set-up. 
The final detection efficiency for 59.5 keV photons obtained for a single PDC in the case of two hit events is $(3\pm 1) \%$.

During laboratory tests of the polarimeter and ensuing comparisons with simulations it became clear that scintillation light was leaking between 
adjacent BGO elements in the PDCs. This was an unexpected effect which will result in a redesign of the PDCs for future flights.  The magnitude 
of the optical cross-talk was mapped using a pulsed LED which was moved between PDCs and the response of all other PDCs was studied. The average 
fraction of light from one PDC reaching a neighbour is 0.95\%, with a range of 0.4\% to 2\% across all PDCs. The resulting optical cross-talk map 
has been integrated into the Geant4 simulation model of the polarimeter.

\section{Monte Carlo simulations}
\label{Geant4}

Tests with radioactive sources cannot reproduce an observation of an astrophysical point source which emits a parallel beam of X-rays with a 
continuous spectrum. The absorption of electromagnetic radiation in the atmosphere also needs to be accounted for.
Monte Carlo simulations can take all these effects into account.

Two different Monte Carlo simulations have been developed.
The first one describes the interactions of particles with the detector array and surrounding materials using the Geant4 package.
The second one simulates how the resulting energy deposits are detected, selected and stored by the instrument. 

The first case depends only on geometry and material choices, and is not affected by instrument settings. 
Since the detection geometry is fixed, this first simulation (which can be very time consuming) needs to be run only once. 
The G4EmLivermorePolarizedPhysics \cite{G4_manual} physics list is used.
The rotation of the polarimeter is simulated in $0.1^\circ$ steps by updating the geometry through a newly developed macro-command. 
To correctly reproduce the 
production of scintillation light, the energies deposited along the path of ejected 
electrons are summed. All data is recorded in ROOT \cite{root} files for a compact storage and 
straight-forward viewing and analysis.

The second Monte Carlo simulation reads the output of the Geant4 simulation 
and translates each energy deposit into scintillation light and a corresponding PMT pulse height. The method described in \cite{MonteCarlo_paper} 
is followed and takes into account scintillator light-yield non-linearity, the attenuation of scintillation light, the production of 
photo-electrons at the PMT photo-cathode, the PMT gain, and finally 
digitisation in units of ADC channels. Results from calibration measurements are used in the
simulation to take into account the deposited energy to photo-electron ratio (i.e. absolute light yield) and the deposited energy to ADC 
channels ratio for each detector unit. 
As the simulation does not produce the PMT waveforms, it is assumed that discrimination between a fast and a slow waveform is not possible below 
100 ADC ($\sim$5 keV). This effect can lead to a fake fast event. Optical cross-talk between detector units is also modelled. 
Using the measured optical cross-talk map as input, for every energy deposit in a PDC, a fraction of energy (equivalent to scintillation light) 
is transferred through-out the detector array to the other PDCs. This procedure is applied just before the 
PMT signal is produced. After conversion into ADC channels, the effect of data acquisition system thresholds and the veto system is considered. 
The simulated veto system uses an approximation of the online and offline selection criteria described in sections 
\ref{Acq_trigger} and \ref{event_sel}, respectively. The exact veto system is not possible to simulate since the simulation 
does not produce waveform signals.

\begin{figure}[!ht]
 \centering
  \includegraphics[width=11 cm]{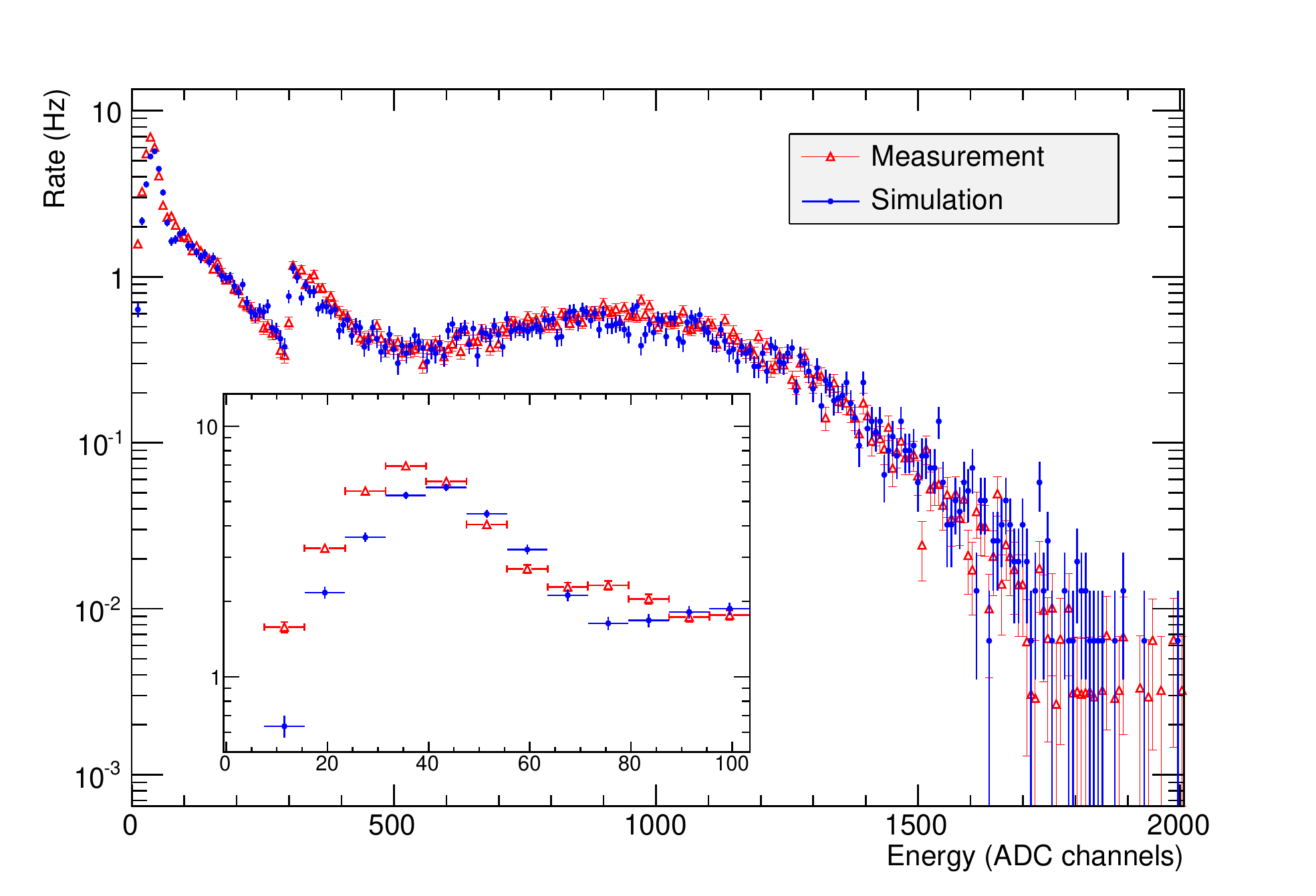}
  \caption{PoGOLite two hit spectrum for a collimated $^{241}$Am source illuminating the central 
PDC. The red (open triangle) spectrum is from calibration, the blue (filled circle) from simulation (no normalisation correction is applied). 
A close-up view shows the low energy part of the spectrum.}
  \label{fig:2hitspectrum}
\end{figure}

\begin{figure}[!ht]
 \centering
  \includegraphics[width=11 cm]{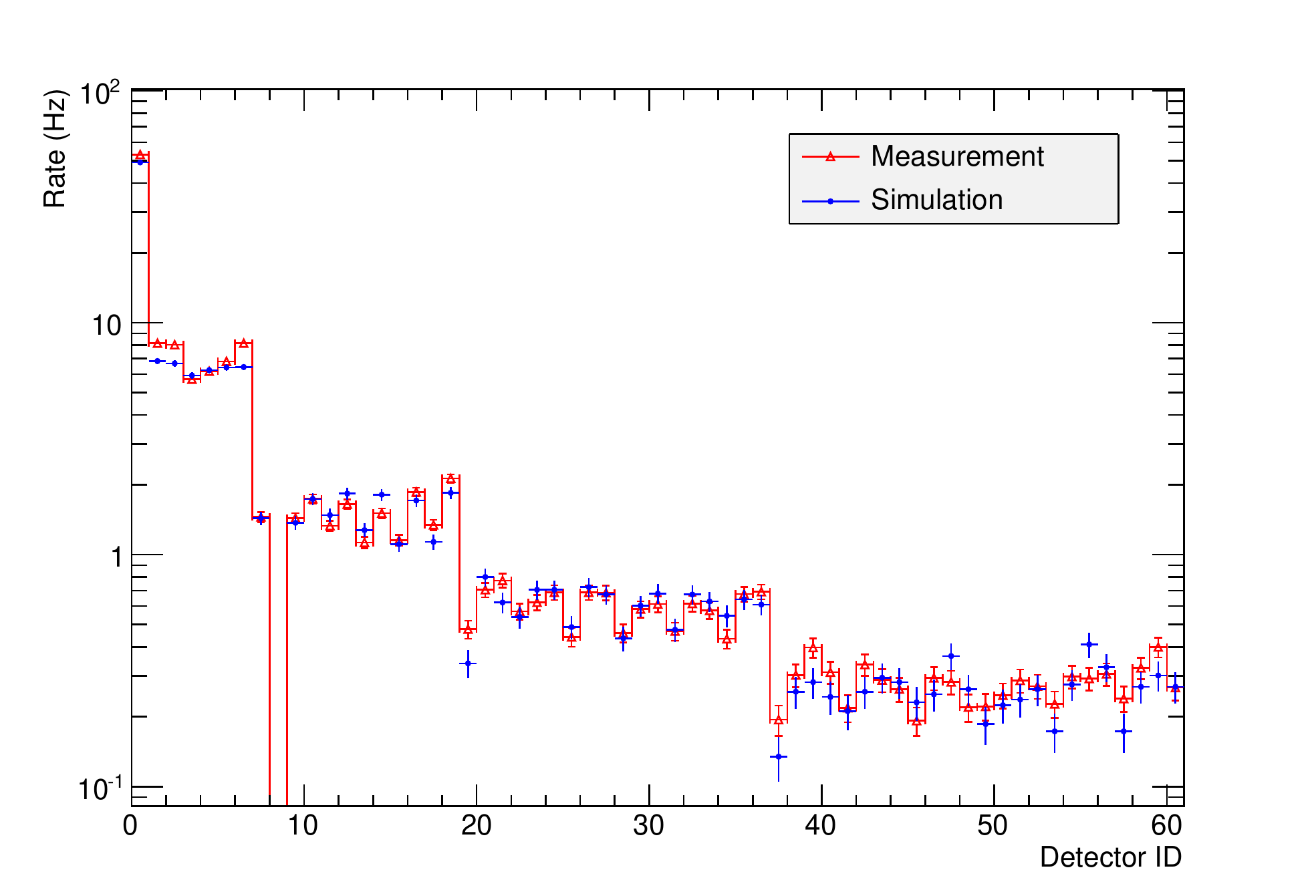}
  \caption{PoGOLite two hit event distribution for a collimated $^{241}$Am source illuminating 
the central PDC. The red (open triangle) distribution is from calibration, the blue (filled circle) from simulation (no normalisation 
correction is applied). 
The detector ID numbering starts at zero for the central PDC and increments in a clockwise spiral towards the peripheral PDCs. 
The gap at detector ID 8 corresponds to a broken PMT which is masked out of the simulation.}
  \label{fig:2hitdistribution}
\end{figure}

To validate the simulations, different configurations have been tested against instrument
measurements. Verification of the energy response, the event distribution
in the detectors and the response to polarisation are presented here. 
Figure \ref{fig:2hitspectrum} and \ref{fig:2hitdistribution} show the two hit energy spectrum and hit
distribution from measurement (red, open triangles) and simulations (blue, filled dots). 
A collimated $^{241}$Am source is placed on top of the central PDC (see the set-up on the left of the figure~\ref{fig:setups}) for these tests. 
The $^{241}$Am emission spectrum measured by a liquid nitrogen cooled germanium detector is used in the simulation, as well as the known source 
rate.
In figure \ref{fig:2hitspectrum} a detailed view of 
the low energy part of the spectrum shows some discrepancies between the measurement and the simulation. The origin of this discrepancy has not 
been understood, but as this region of the spectrum corresponds to single photo-electron emission it is likely to be related to the PMT behaviour. 
Above 100 ADC channels there is good agreement between measurement and simulations.
Figure \ref{fig:2hitdistribution} also shows good agreement between measurement
and simulations as no normalisation corrections are applied between the two distributions. The variation in number of events between detectors 
is closely related to the detector array geometry, differences in PDC sensitivity and optical cross talk. The good agreement shows that the instrument
response is well reproduced in the simulations.

\section{Instrument response to polarisation}
\label{inst_pol_rsp}

The polarimetric response of the instrument is determined with measurements and verified with simulations. It is important to study the response 
to unpolarised beams and backgrounds and identify potential systematic effects.
In flight backgrounds may be anisotropic, e.g. due to albedo atmospheric neutrons resulting in more events in the Earth-facing part of the 
polarimeter.

\begin{figure}[!ht]
 \centering
\includegraphics[width=11 cm]{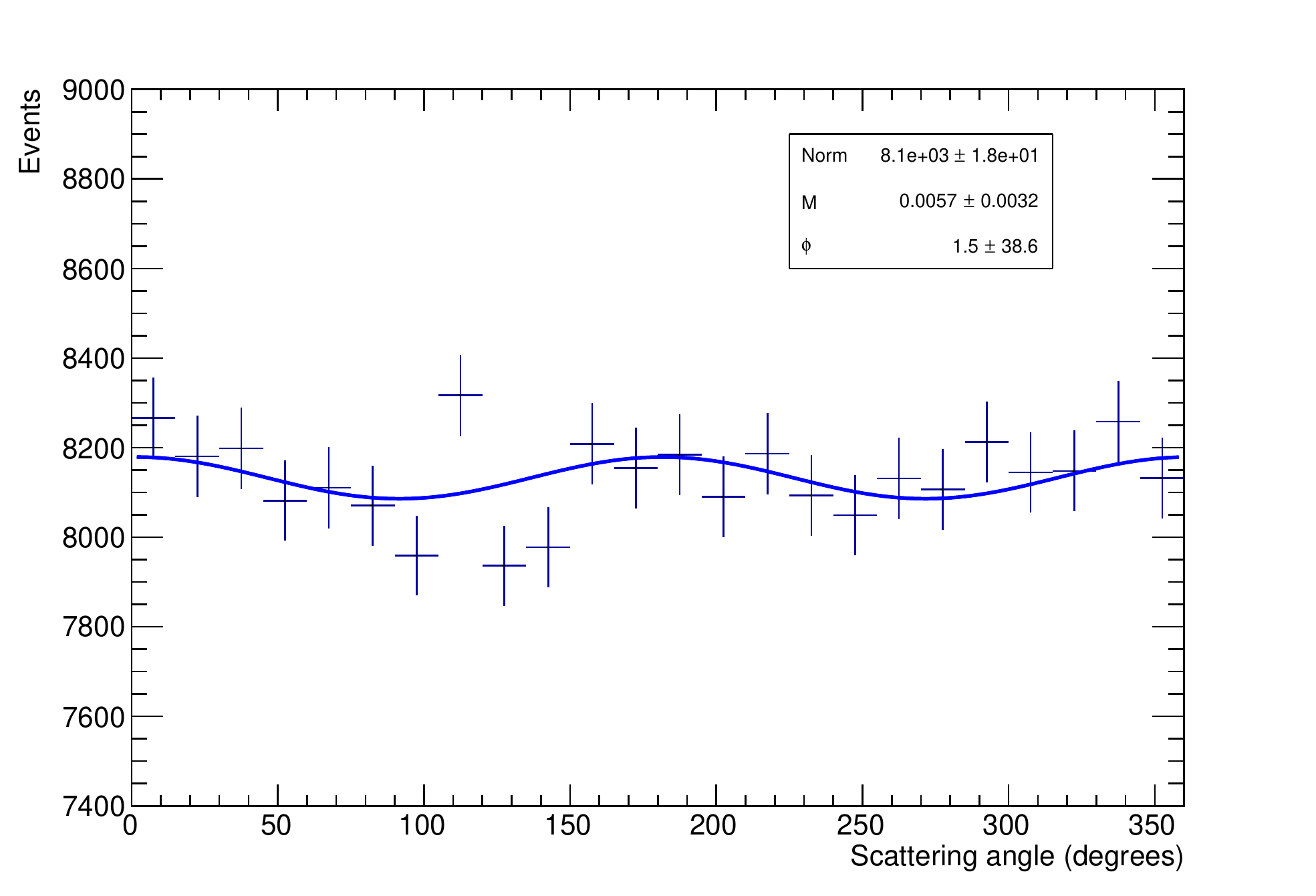}
\caption{Scattering angle distribution for an unpolarised beam of $^{241}$Am photons fitted with Equation \ref{modul_factor}. }
\label{fig:unpol}
\end{figure}

The $^{241}$Am source was placed on top of the central PDC unit (see the set-up on the left side of figure~\ref{fig:setups}) and the whole 
detector array was rotated, making a full rotation every 5 minutes. The resulting modulation curve is shown in 
figure \ref{fig:unpol}. The source rate is several orders of magnitude higher than the room background rate.
The resulting modulation factor is $(0.57\pm 0.32)\%$ (fit $\chi^{2} /dof = 21.9/21 $) which is 1.8$\sigma$ from zero modulation. 
Since measurement of a low polarisation fraction does not follow Gaussian statistics, the Rice distribution as in \cite{Strohmayer} was used to 
determine that the $99\%$ upper limit on the modulation factor is $1.38\%$.
A simulation of the set-up yielded a modulation factor of $(0.51\pm0.38)\%$ which is fully compatible with the measured result.

\begin{figure}[!ht]
 \centering
\includegraphics[width=11 cm]{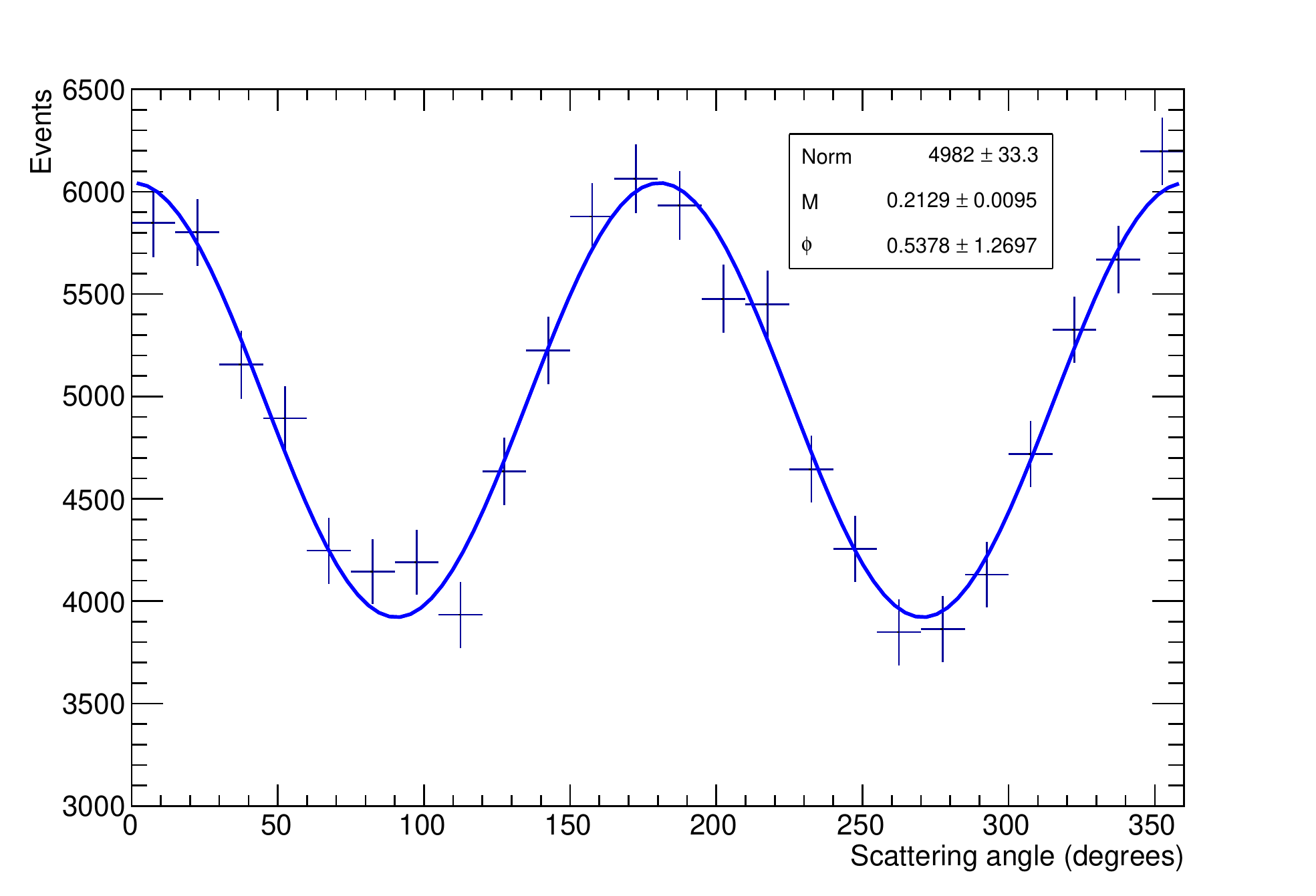}
\caption{Scattering angle distribution for a polarised beam of $^{241}$Am photons fitted with Equation \ref{modul_factor}.}
\label{fig:polAm}
\end{figure}

To create a polarised beam the 59.5 keV photons from the $^{241}$Am source were scattered through $90^{\circ}$ on a piece of 
polyethylene resulting in a nearly 100\% polarised beam of 53.3 keV photons. The exit aperture of the scatterer was aligned with the central PDC
(see the set-up on the right side of the figure~\ref{fig:setups}). 
Because the room background rate is comparable to the rate of polarised photons, the final modulation curve needs to be background subtracted. 
The modulation curve after background subtraction is shown in figure \ref{fig:polAm}. The measured modulation factor is
$M_{\textrm{meas}} = (21.3\pm0.9)\%$ (fit $\chi^{2} /dof = 17.8 / 21 $). 
Simulations yielded a modulation factor, $M_{\textrm{sim}}$, of $(23.7\pm0.14)\%$ which is compatible within 2 $\sigma$ of the 
measured one. The discrepancy may arise because of simplifications of the event selection in the simulation or it may
reflect the disagreement in the low energy region, as presented in figure \ref{fig:2hitspectrum}. 

\begin{figure}[!ht]
 \centering
\includegraphics[width=11 cm]{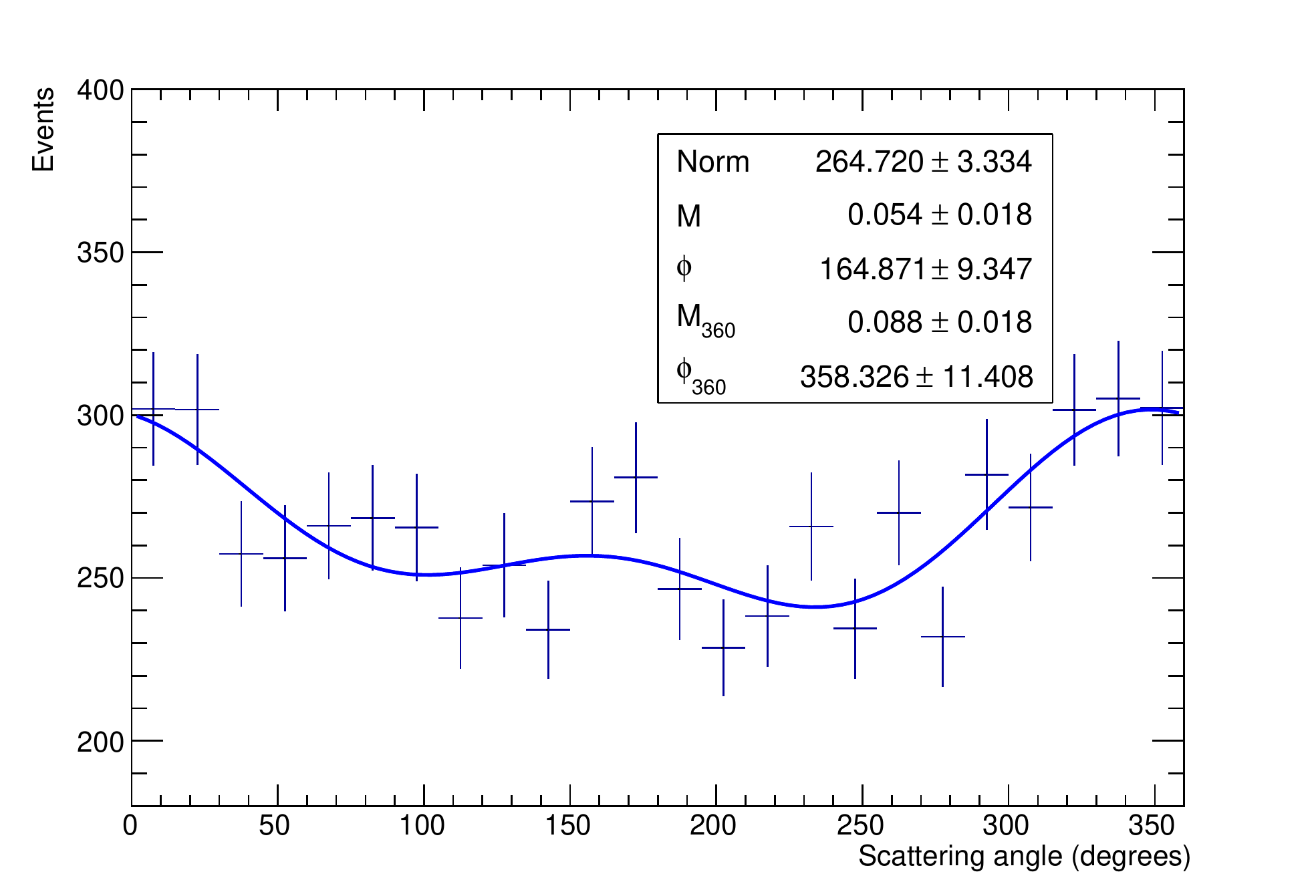}
\caption{Scattering angle distribution for a $^{137}$Cs beam impinging on the side of the polarimeter fitted with Equation 
\ref{modul_factor_180_360}.}
\label{fig:polCs}
\end{figure}

To investigate the effect of an anisotropic background the instrument was irradiated from the side with a $^{137}$Cs source. The resulting 
modulation curve, shown in figure \ref{fig:polCs}, presents two contributing components with $360^{\circ}$ and $180^{\circ}$ periodicity.
At energies much greater than $100\,\textrm{keV}$ (as is the case for $^{137}$Cs) a photon has a high probability to Compton scatter twice in 
the detector rather than undergoing photo-absorption. As it is not possible to determine which scattering occured first, the direction of the 
photon cannot be reconstructed, leading to a  $180^{\circ}$ component in the 
modulation curve. If the primary and secondary interactions can be distinguished, a $360^{\circ}$ component
is created - as in the measured case where this component has a 4.9 $\sigma$ significance (fit $\chi^{2} /dof = 52/55$). The rotation of the 
instrument does not remove such modulations as they are external to the detector array. This result shows the importance of measuring 
background anisotropies during flight. Resulting background modulation curves may then be subtracted from those derived from source observations.

\section{Scientific performance}
\label{performance}

\begin{figure}[!ht]
  \centering
    \includegraphics[width=11 cm]{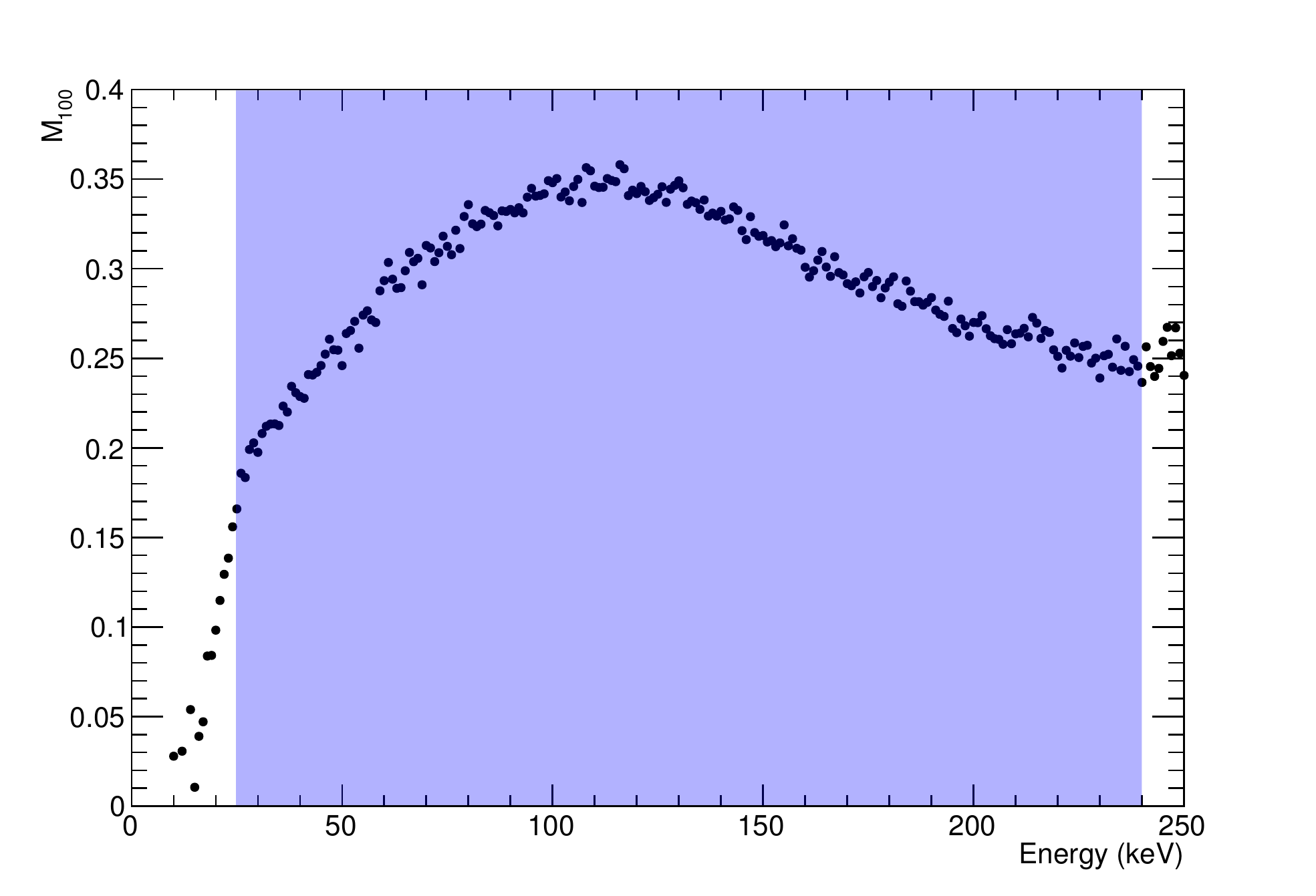}
\caption{The simulated value of $M_{100}$ as a function of energy. The shaded area represents the PoGOLite energy range. }
 \label{fig:M100_energy}
\end{figure}

\begin{figure}[!ht]
 \centering
  \includegraphics[width=11 cm]{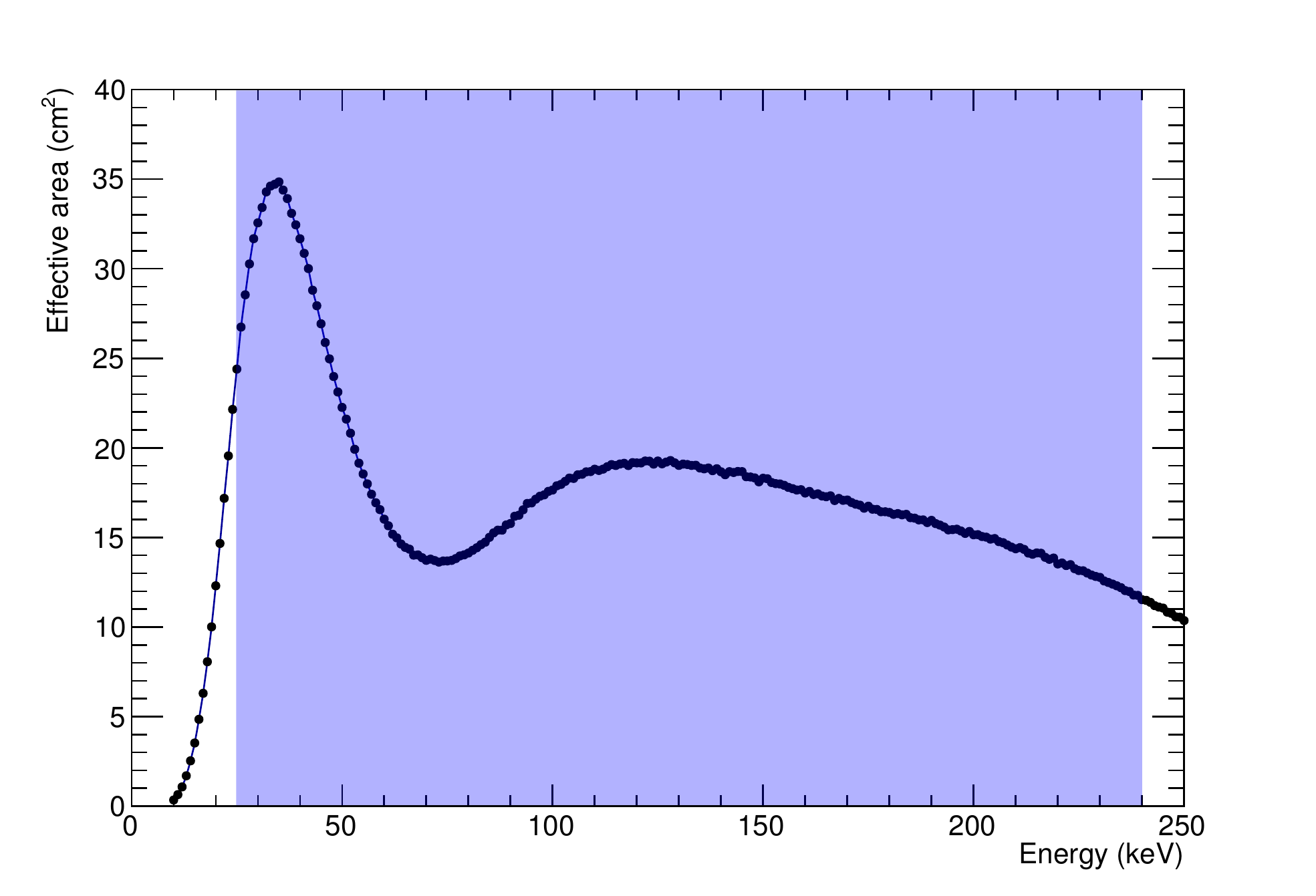}
  \caption{Instrument effective area as a function of energy. For exceedingly low energies, the Compton scattering becomes indistinguishable
  from the read-out noise level, while for high energies, the cross-section for a photo-absorption diminishes. The first peak corresponds to 
  events with one Compton scattering and one photo-absorption, while the second peak, around \mbox{120 keV}, is due to Compton-Compton events 
  where the third photon interaction is not detected or not in the detector area. The energy range of the instrument has been indicated.}
    \label{fig:Eff_area}
\end{figure}

After showing the simulation results to be compatible with measurements, a series of Geant4 Monte Carlo simulations were performed to study 
the scientific potential of PoGOLite observations of the Crab. The 
polarimeter was illuminated by a mono-energetic fully polarised photon beam for the energy range 10 keV to 
240 keV. The instrumental energy range, 25-240 keV, is determined by the lowest photon energy that can trigger the data acquisition system 
and by the sum of the maximum reconstructed energy per single hit in the case of two hit events. The energy dependence of the effective area of the 
polarimeter and $M_{100}$ are presented in figure \ref{fig:M100_energy} and figure \ref{fig:Eff_area}, respectively. 

\begin{figure}[!ht]
  \centering
    \includegraphics[width=11 cm]{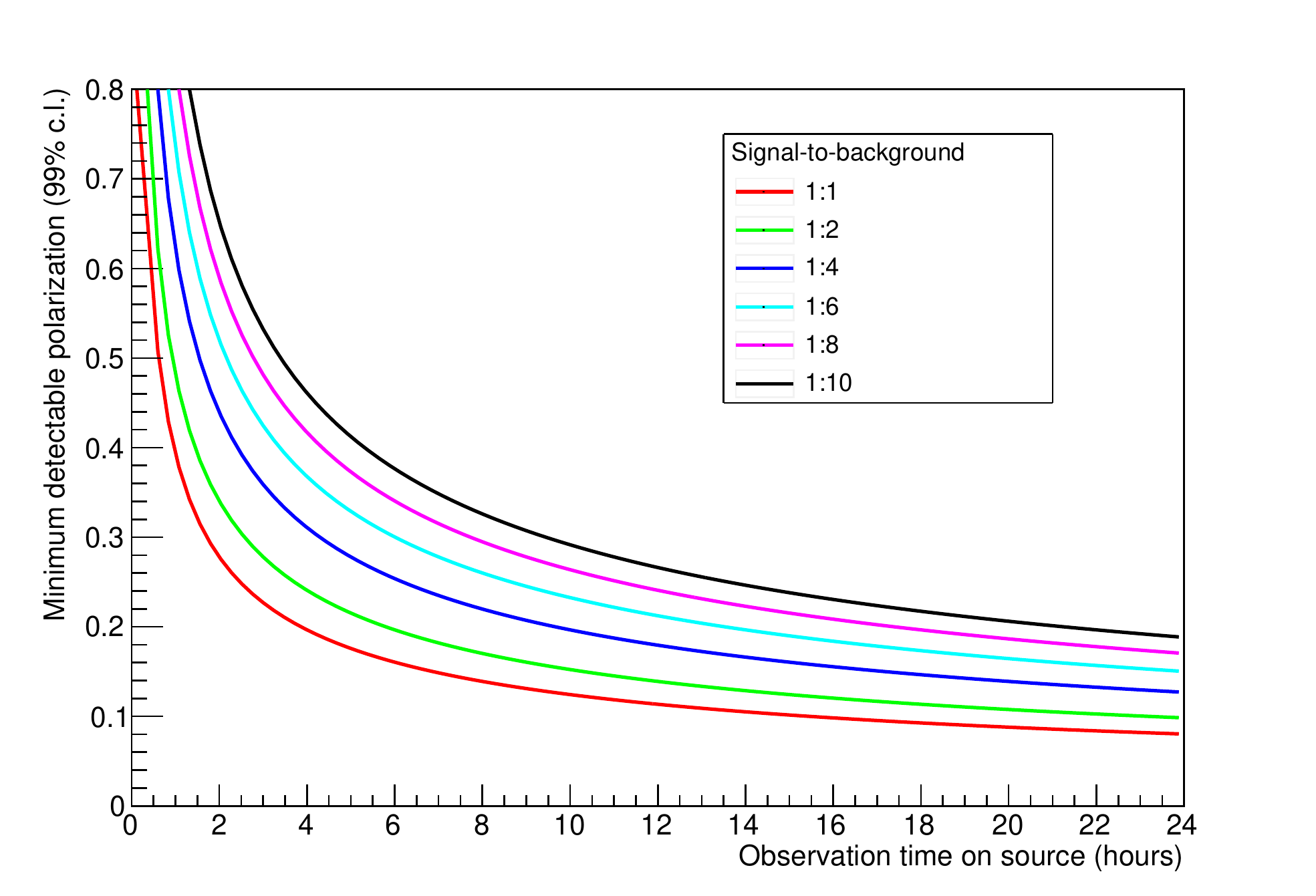}
\caption{The MDP as a function of observation time for different scenarios of signal-to-background ratio.}
 \label{fig:MDP_bkg}
\end{figure}

A simulation was performed to determine $M_{100}$ for a Crab observation with an atmospheric path length of $5\,\mathrm{g/cm^2}$, as expected for 
this balloon flight.
The X-ray flux from the Crab in the hard X-ray band can be approximated as $F(E)=9.7E^{-2.1}$ \cite{Toor}. It is noted that the residual
atmosphere above the polarimeter significantly suppresses the flux below $\sim$20 keV. 
The reconstructed modulation factor was $M_{\textrm{Crab, sim}}=(24.85\pm0.15)\%$. 
The modulation factor for the Crab is given by 
$({M_{\textrm{meas}}} / {M_{\textrm{sim}}}) \times M_{\textrm{Crab, sim}}=(22.3\pm1.2)\%$, where the
scaling factor has been applied to take into account the discrepancy between simulation and measurement reported in section~\ref{inst_pol_rsp}.
Using these results the achievable MDP for Crab observations can be evaluated for different signal-to-background conditions, as shown in figure 
\ref{fig:MDP_bkg}. 
The in-flight signal rate of 1.5 Hz has been evaluated using the Geant4 simulation.

\section{Conclusions}
\label{conclusion}

PoGOLite is a balloon-borne polarimeter which operates in the hard X-ray band, 25 to 240 keV.  The narrow field-of-view, 
$2.4^\circ \times 2.6^\circ$, defined by an active and passive collimator system is necessary for the polarimeter to be optimised for the 
observation of point sources. 
A veto system allows background events to be identified and rejected. This reduces dead-time without significantly reducing the acceptance 
of signal events. A maximum signal rate of 1.25~kHz can be correctly reconstructed which is significantly higher than the rate expected in-flight. 
Signal events are time-tagged with an accuracy of order 1~$\mu$s which makes the polarimeter suitable for the study of the Crab pulsar. 
Offline selections made on downlinked data allow more stringent constraints to be placed on data resulting in only 5\% of slow scintillator 
waveforms leading to the generation of a trigger signal. Over 90\% of fast scintillator hits are identified as signal.
Tests have demonstrated that the BGO anticoincidence shield is not 100\% efficient - high energy photons are able to pass into the sensitive 
volume of the polarimeter without a veto signal being issued. For this reason, background conditions must be carefully studied during flight 
and source modulation curves corrected accordingly.
A detailed Geant4 simulation of the polarimeter has been set-up and calibrated with data collected during tests with radioactive sources. The 
response of the polarimeter is reproduced very well except at lower part of the energy range, below 100 ADC ($\sim$5 keV).
The response of the polarimeter has been studied using an unpolarised beam of photons. The reconstructed modulation factor is compatible 
with zero. 
The $M_{100}$ for a Crab observation is simulated to be $\sim$22\%. For a single 6 hour long observation 
of the Crab the predicted MDP ranges between $\sim$16\% and $\sim$38\% for a signal-to-background of 1:1 and 1:10, respectively.
The $99\%$ upper limit on the unpolarised modulation factor (presented in section~\ref{inst_pol_rsp}) scaled with M100 is $6.17\%$. 
This is below the MDP predictions, as required for a meaningful measurement.

\section{Acknowledgements}

The PoGOLite Collaboration acknowledges funding from The Swedish National Space Board, The Swedish Research Council, The Knut and Alice Wallenberg 
Foundation. Stefan Rydstr\"{o}m is thanked for his significant contributions to the design and construction of the polarimeter and for his help in 
setting up tests. DST Control developed the PoGOLite attitude control system which was used in some of the measurements reported here. 
The SSC Esrange Space Centre is thanked for their hospitality during the pre-flight test campaign. Tsunefumi Mizuno is thanked for his advice 
regarding the Geant4 simulations. Torbj\"{o}rn B\"{a}ck made important contributions to the characterisation of the $^{241}$Am source.


\bibliographystyle{elsarticle-num}



\end{document}